\documentclass[aps,prd,preprint,longbibliography,floatfix]{revtex4-1}
\usepackage{amsmath, amsthm, amssymb, graphicx}
\usepackage{slashed}
\usepackage{physics}
\usepackage{xcolor}
\usepackage[colorlinks=true,linkcolor=blue,citecolor=blue,urlcolor=blue]{hyperref}

\begin{document}

% Use the \preprint command to place your local institutional report number in the upper righthand corner of the title page in preprint mode. Multiple \preprint commands are allowed. Use the 'preprintnumbers' class option to override journal defaults to display numbers if necessary
%\preprint{}

% Title of paper
%\title{Microscopic production of gravitational wave  from particles produced by bubble collisions during cosmological FOPT}

%\title{Associated gravitational wave production from bubble collision during a first-order phase transition}

\title{High-frequency gravitational waves from microscopic particle-graviton associated production in bubble collisions}

% repeat the \author .. \affiliation  etc. as needed \email, \thanks, \homepage, \altaffiliation all apply to the current author. Explanatory text should go in the []'s, actual e-mail address or url should go in the {}'s for \email and \homepage. Please use the appropriate macro foreach each type of information

% \affiliation command applies to all authors since the last \affiliation command. The \affiliation command should follow the other information \affiliation can be followed by \email, \homepage, \thanks as well.
\author{Peilin Chen}
\author{Dayun Qiu}
\author{Zihong Cheng}
\author{Fa Peng Huang}
\email{Corresponding Author. huangfp8@sysu.edu.cn}
\affiliation{%
MOE Key Laboratory of TianQin Mission, TianQin Research Center for Gravitational Physics \& School of Physics and Astronomy, Frontiers Science Center for TianQin, Gravitational Wave Research Center of CNSA, Sun Yat-sen University (Zhuhai Campus), Zhuhai 519082, China
}%

%\date{\today}

% Collaboration name if desired (requires use of superscriptaddress option in \documentclass). \noaffiliation is required (may also be used with the \author command). \collaboration can be followed by \email, \homepage, \thanks as well.
%\collaboration{}
%\noaffiliation

\begin{abstract}
Cosmological first-order phase transitions are well-known sources of gravitational waves from bubble collisions and plasma dynamics. In this work, we investigate an additional microscopic source directly arising from the associated production of particles and gravitons during bubble collisions. Focusing on runaway bubble walls, we calculate the gravitational wave spectra from two-body scalar–graviton production and three-body processes involving scalar or fermion pairs accompanied by a graviton. We derive analytic approximations and numerically evaluate the resulting spectra, revealing a high-frequency gravitational wave component that can complement conventional lower-frequency signals from the same phase transition. This mechanism offers a potential multiband signature and establishes a connection between gravitational radiation and the production of dark matter or other new particles. Our findings provide a new physics target for future high-frequency gravitational wave experiments and a complementary probe of particle production in the early Universe.
\end{abstract}

% insert suggested keywords - APS authors don't need to do this \keywords{}

% \maketitle must follow title, authors, abstract, and keywords
\maketitle

\section{Introduction}\label{sec:introduction}

Cosmological first-order phase transitions (FOPTs) provide a promising window into the thermal history of the early Universe and physics beyond the Standard Model. The violent nonequilibrium dynamics associated with bubble nucleation, expansion, and collisions can lead to efficient particle production and generate a stochastic gravitational wave (GW) background~\cite{Watkins:1991zt,Mansour:2023fwj,Shakya:2023pp,Caprini:2015zlo,Caprini:2019egz}. In the conventional picture, the released vacuum energy is converted into scalar field gradients and bulk plasma motion, giving rise to three well-known GW sources: bubble wall collisions, long-lived sound waves, and magnetohydrodynamic turbulence~\cite{Kosowsky:1991ua,Kosowsky:1992vn,Kamionkowski:1993fg,Caprini:2015zlo,Caprini:2019egz}. Extensive numerical simulations have established the importance of acoustic GW production and characterized its spectral properties, while numerical studies of magnetohydrodynamic turbulence have provided further insight into its contribution to the stochastic GW background~\cite{Hindmarsh:2013xza,Hindmarsh:2015qta,Hindmarsh:2017gnf,RoperPol:2019wvy}. These macroscopic sources have been widely investigated as potential signals for future GW observatories. Nevertheless, the highly nonequilibrium environment of a FOPT can also induce microscopic particle processes that provide additional channels for GW generation.

Recent studies have extended the conventional picture by investigating GW sources associated with particle production and propagation during FOPTs. Jinno, Shakya, and van de Vis ~\cite{Jinno:2022fom} considered phase transitions in which a substantial fraction of the released energy is transferred to feebly interacting particles. Instead of forming a strongly interacting fluid, these particles propagate freely, and the anisotropic stress of the resulting particle shells generates a broad GW spectrum with characteristic features distinct from those of acoustic waves. Other studies have explored gravitational radiation directly associated with microscopic particle interactions with bubble walls. In particular, 
%Ai~\cite{Ai:2025fqw}, Qiu, Jiang, and Huang 
Refs.~\cite{Ai:2025fqw,Qiu:2025tmn} studied gravitational bremsstrahlung from heavy particles decelerated by bubble walls, finding that the characteristic frequency depends on the wall velocity and that the GW amplitude exhibits a strong dependence on the heavy particle mass. The heavy particle could undergo decay process even after acquiring mass in the true vacuum.  Ref.~\cite{Bhandari:2026uqd} investigates the GW signals generated by this process.
These developments suggest that microscopic particle processes can produce GW signatures complementary to the conventional macroscopic sources.

Particularly favorable conditions for nonequilibrium particle production can arise during sufficiently strong FOPTs, in which the released vacuum energy accelerates bubble walls to ultrarelativistic velocities. The wall dynamics is governed by the competition between the driving pressure from the vacuum energy difference and the friction exerted by the surrounding plasma~\cite{Bodeker:2009qy,Espinosa:2010hh}. Additional friction from transition radiation can become important at large wall Lorentz factors and may modify or prevent runaway acceleration~\cite{Bodeker:2017cim}. When the bubble walls nevertheless reach highly relativistic velocities and retain a sizable fraction of the released energy, their collisions generate rapidly varying and spatially inhomogeneous scalar field configurations. The resulting classical background provides a source of energy and momentum for the production of particles that may otherwise be difficult to produce in a thermal environment~\cite{Falkowski:2012fb,Mansour:2023fwj,Giudice:2024pp,Lewicki:2019gmv}. Such processes are particularly interesting for the production of dark matter and other new particles beyond the Standard Model~\cite{Katz:2016adq,Mansour:2023fwj,Shakya:2023pp,Giudice:2024pp,Ai:2024ikj,Cataldi:2025nac,Cheng:2026npt,Ghoshal:2026hev}.

The particles produced during bubble collisions can themselves generate gravitational radiation. Inomata, Kamionkowski, Kasai, and Shakya investigated GW production from the inhomogeneous particle distributions left behind after bubble collisions. The anisotropic stress of these distributions persists after the bubble walls disappear and can modify the resulting GW spectrum, including its low-frequency behavior~\cite{Inomata:2024rkt}. However, gravitational radiation need not arise exclusively from the subsequent evolution of the produced particles. Since particles couple universally to gravity, gravitons can also be emitted directly during the microscopic particle production processes induced by the collision background. This possibility motivates the investigation of an additional GW source that is generated at the moment of particle production, rather than by the later evolution of the resulting particle distribution. In this work, we investigate a distinct microscopic source of GWs
arising from graviton emission accompanying particle production
during bubble collisions. Unlike gravitational radiation sourced
by the subsequent propagation of produced particles or by
particle interactions with individual bubble walls, the mechanism
considered here originates directly from particle--graviton
associated production induced by the rapidly varying collision
background. Focusing on the runaway-wall regime, we first
investigate the two-body channel
\begin{equation}
    \phi(l)\rightarrow \phi(p)+h_{\mu\nu}(k),
\end{equation}
where $\phi(l)$ denotes a Fourier component of the classical
scalar background generated by the bubble collisions, rather
than an incoming physical particle. This background supplies
the energy and momentum required to produce an on-shell scalar
particle $\phi(p)$ and a graviton $h_{\mu\nu}(k)$.
We further consider the three-body channels
\begin{align}
    \phi(l)&\rightarrow
    \varphi(p)+\varphi^\ast(q)+h_{\mu\nu}(k),
    \\
    \phi(l)&\rightarrow
    \psi(p)+\bar{\psi}(q)+h_{\mu\nu}(k),
\end{align}
in which scalar or fermion particle--antiparticle pairs are
produced together with a graviton. Depending on their masses
and interactions, the additional scalar and fermion fields
may represent dark matter candidates or other new particles
beyond the Standard Model. These processes establish a direct
connection between nonthermal particle production and
gravitational radiation during bubble collisions.
To quantify this new GW source, we formulate the particle
production rates in terms of the collision efficiency factor
and the imaginary part of the two-point one-particle-irreducible
Green function. We calculate the differential graviton spectra
for both the two-body and three-body channels and derive
analytic approximations that elucidate their characteristic
spectral behavior. Combining these analytic results with
numerical calculations, we determine the GW energy spectra
at production and obtain their present-day spectra by
accounting for cosmological redshift. In particular, we
investigate the dependence of the resulting spectra on
the bubble-wall Lorentz factor and the masses of the
produced particles, with an emphasis on their high-frequency
behavior.
An important motivation for studying this mechanism is its
potential to generate high-frequency GW signals complementary
to those from conventional phase-transition sources.
Whereas the characteristic frequencies of the latter
are primarily determined by macroscopic scales associated
with bubble expansion and plasma dynamics, microscopic
particle--graviton production can probe the considerably
higher energy scales accessible during relativistic bubble
collisions. A single FOPT may therefore generate GW signals
in widely separated frequency bands, providing a potential
multiband observational signature. Furthermore, the
associated production of gravitons and dark matter or
other new particles offers a complementary means of
probing the microscopic interactions responsible for
nonthermal particle production. The resulting high-frequency
GW component provides an additional science target for
future high-frequency GW experiments and may help uncover
particle physics beyond the Standard Model that is
inaccessible through conventional GW observations alone. 

The remainder of this paper is organized as follows.
In Sec.~\ref{sec:collision_framework}, we introduce the
theoretical framework for particle production from
bubble collisions and the formalism for calculating
the associated graviton emission.
In Sec.~\ref{sec:two_body_emission}, we calculate the
differential graviton spectrum for the two-body scalar
channel, derive analytic approximations, and present
the corresponding numerical results and present-day
GW spectrum.
In Sec.~\ref{sec:three_body_emission}, we extend the
analysis to three-body processes involving scalar
and fermion pairs and investigate their respective
GW spectra. Finally, we summarize our main findings in
Sec.~\ref{sec:conclusion}.

%The remainder of this paper is organized as follows. In Sec.~\ref{sec:collision_framework}, we introduce the formalism for calculating particle production from bubble collisions. Secs.~\ref{sec:two_body_emission} and~\ref{sec:three_body_emission} present the calculations and present-day spectra for the two-body scalar channel and the three-body scalar and fermion channels, respectively. The numerical power spectrum of each channel is displayed separately in the corresponding section. Sec.~\ref{sec:motivation} briefly summarizes the motivation for this microscopic source, and our conclusions are given in Sec.~\ref{sec:conclusion}.

\section{Graviton emission from bubble collisions}
\label{sec:collision_framework}

A strong FOPT releases vacuum energy and, when friction exerted by the surrounding plasma is sufficiently weak, accelerates the bubble walls to ultrarelativistic velocities. In the runaway regime, collisions between expanding bubbles generate rapidly varying and highly inhomogeneous classical scalar field configurations in the collision region, which can efficiently produce particles~\cite{Falkowski:2012fb,Mansour:2023fwj,Giudice:2024pp,Lewicki:2019gmv}.
In this work, we investigate the associated production of gravitons together with scalar or fermionic particles, and show that these processes can provide a distinct microscopic source of GWs.

%This excited background can produce physical particles together with gravitons, providing a microscopic contribution to the stochastic GW background. The process is illustrated schematically in Fig.~\ref{fig:bubble_collision_particle_graviton_channels}.

\begin{figure*}[h]
\centering
\includegraphics[width=0.98\textwidth]{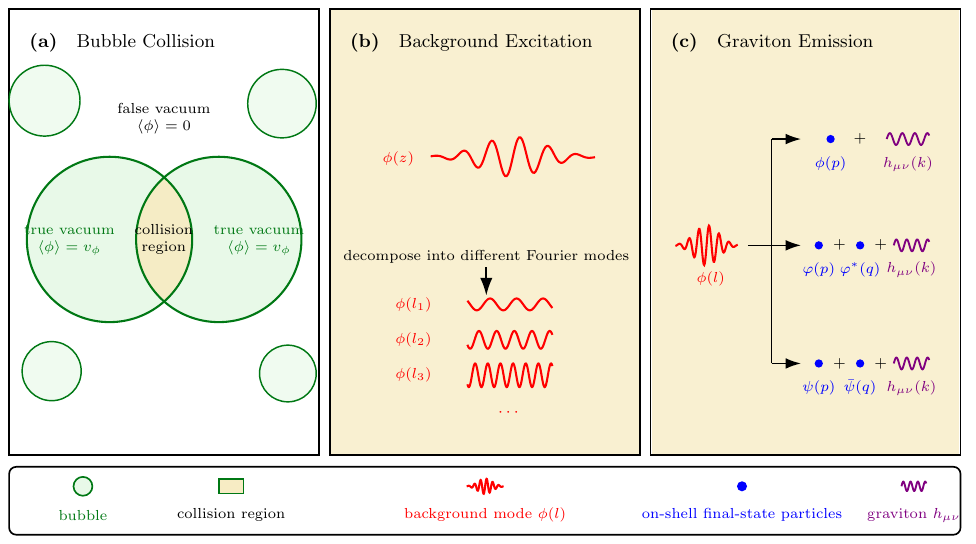}
\caption{Schematic illustration of graviton emission associated with particle production during bubble collisions. (a) Two expanding true-vacuum bubbles collide and generate a highly excited collision region. (b) The collision generates a classical background excitation, which is decomposed into Fourier components $\phi(l_i)$. (c) These background modes supply energy and momentum to the physical final states in the two-body channel $\phi(l)\to\phi(p)+h_{\mu\nu}(k)$ and in the scalar and fermion three-body channels $\phi(l)\to\varphi(p)+\varphi^\ast(q)+h_{\mu\nu}(k)$ and $\phi(l)\to\psi(p)+\bar\psi(q)+h_{\mu\nu}(k)$.}
\label{fig:bubble_collision_particle_graviton_channels}
\end{figure*}

The microscopic origin of this new GW source considered here is summarized schematically in Fig.~\ref{fig:bubble_collision_particle_graviton_channels}. Panel~(a) depicts the collision of two expanding true-vacuum bubbles and the formation of a highly excited collision region. The resulting classical field configuration is represented in panel~(b) through its Fourier decomposition into modes $\phi(l_i)$. These modes encode the energy and momentum carried by the time-dependent collision background and should not be regarded as physical incoming particles. Panel~(c) displays the particle-production channels studied in this work. A background mode can source a physical excitation of the same scalar field accompanied by graviton emission,
$\phi(l)\to\phi(p)+h_{\mu\nu}(k)$.
If the background field couples to additional scalar or fermionic degrees of freedom, graviton-associated three-body production,
$\phi(l)\to\varphi(p)+\varphi^\ast(q)+h_{\mu\nu}(k)$
and
$\phi(l)\to\psi(p)+\bar\psi(q)+h_{\mu\nu}(k)$,
can also occur. In each case, $\phi(l)$ denotes a Fourier mode of the classical collision background that injects the total four-momentum into the physical final state, rather than an asymptotic incoming particle.

The Fourier component of the collision background \(\phi(l)\) carries four-momentum
\begin{equation}
    l^\mu=(E,\boldsymbol{\ell})=(E,0,0,\ell),
    \qquad
    \chi\equiv l_\mu l^\mu=E^2-\ell^2.
    \label{eq:collision_background_virtuality}
\end{equation}
Here \(E\) and \(\ell\) are the background energy and longitudinal momentum variables in the collision frame, and \(\chi\) is the invariant four-momentum squared. The background component \(\phi(l)\) can carry values of \(\chi\) that are not fixed by the scalar mass \(m_\phi\).

The number \(N\) of emitted gravitons per unit collision area~\cite{Watkins:1991zt,Falkowski:2012fb,Mansour:2023fwj,Giudice:2024pp},
\begin{equation}
    \frac{N}{A}
    =
    \int_{\Lambda_{\min}}^{\Lambda_{\max}} dE
    \int_{\Lambda_{\min}}^{\Lambda_{\max}} d\ell\,
    f_{\mathrm{coll}}(E,\ell)
    \operatorname{Im}\!\left[\widetilde{\Gamma}^{(2)}(E,\ell)\right].
    \label{eq:graviton_yield_per_collision_area}
\end{equation}
The two factors in Eq.~\eqref{eq:graviton_yield_per_collision_area} have distinct physical roles. The collision efficiency factor \(f_{\mathrm{coll}}(E,\ell)\) measures the spectral weight supplied by the evolving walls at energy \(E\) and longitudinal momentum \(\ell\). 
The imaginary part \(\operatorname{Im}[\widetilde{\Gamma}^{(2)}(E,\ell)]\) describes the production of physical final states by a background component with four-momentum \(l^\mu\). Here \(\widetilde{\Gamma}^{(2)}\) is the Fourier-transformed two-point one-particle-irreducible (1PI) function, and only contributions with one emitted graviton are retained in Eq.~\eqref{eq:graviton_yield_per_collision_area}. 

The lower and upper cutoffs are associated with the inverse characteristic bubble radius \(R_*\) and the inverse Lorentz-contracted wall thickness \(l_{\mathrm{wall}}\), respectively. For the numerical calculation, we impose the definite cutoffs
\begin{equation}
\Lambda_{\min}\equiv R_*^{-1},
\qquad
\Lambda_{\max}\equiv \gamma_{\mathrm{wall}}v_\phi,
\label{eq:collision_background_ir_uv_cutoffs}
\end{equation}
where the prescription for \(\Lambda_{\max}\) is based on the order-of-magnitude estimate \(l_{\mathrm{wall}}^{-1}\sim\gamma_{\mathrm{wall}}v_\phi\). The corresponding phase transition parameters and the validity conditions of the planar approximation are discussed in Appendix~\ref{app:particle_production_formalism}.

For the elastic ultrarelativistic thin-wall profile used here, including the background cutoffs specified above, we use
\begin{equation}
    f_{\mathrm{coll}}(E,\ell)=\frac{32v_\phi^2}{\pi^2\chi^2}\Theta(E-\Lambda_{\min})\Theta(\Lambda_{\max}-E)\Theta(\ell-\Lambda_{\min})\Theta(\Lambda_{\max}-\ell).
\label{eq:thin_wall_collision_efficiency}
\end{equation}
It therefore describes which background components are available to produce particles and how strongly they contribute.

Through the optical theorem, this imaginary part is expressed as a sum over the corresponding production channels,
\begin{align}
    \operatorname{Im}\!\left[\widetilde{\Gamma}^{(2)}(E,\ell)\right]
    &=
    \frac12\sum_c\int d\Pi_c\,
    \overline{|\mathcal M_c|^2}
    \Theta\!\left[\chi-\chi_{\mathrm{th}(c)}\right],
    \label{eq:optical_theorem_channel_sum}
    \\
    d\Pi_c
    &=
    (2\pi)^4\delta^{(4)}\!\left(l-\sum_i k_i\right)
    \prod_i\frac{d^3\boldsymbol{k}_i}{(2\pi)^3\,2E_i}.
    \label{eq:final_state_lorentz_phase_space}
\end{align}
Here \(\overline{|\mathcal M_c|^2}\) is the squared transition amplitude for channel \(c\), summed over final state spins and graviton polarizations. It contains the couplings and masses that determine the strength and momentum dependence of the production process. The Lorentz-invariant phase-space measure \(d\Pi_c\) sums over the allowed momenta of the physical final-state particles. Its delta function enforces energy and momentum conservation, and the energies \(E_i\) satisfy the corresponding on-shell conditions. The threshold \(\chi_{\mathrm{th}(c)}\) is the minimum invariant four-momentum squared required to produce channel \(c\), so the step function excludes configurations below that threshold. Since each selected channel contains one graviton, each production event contributes one to \(N\).

Thus, Eq.~\eqref{eq:graviton_yield_per_collision_area} combines the energy and momentum content of the collision background with its conversion into particles accompanied by graviton emission. Although the inclusive phase-space integral is Lorentz invariant and can be evaluated in the center-of-momentum frame, the differential graviton energy spectrum depends on the reference frame. We therefore retain the collision frame variables \((E,\ell)\) in the following calculations. A derivation of Eq.~\eqref{eq:graviton_yield_per_collision_area}, including the planar normalization and the relation between the wall profile and \(f_{\mathrm{coll}}\), is given in Appendix~\ref{app:particle_production_formalism}.

\section{Two-body production with graviton emission}
\label{sec:two_body_emission}

\begin{figure}[h]
    \centering
    \includegraphics[width=0.72\columnwidth]{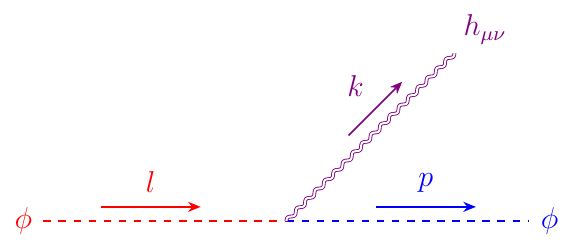}
    \caption{Tree-level two-body process $\phi(l)\to\phi(p)+h_{\mu\nu}(k)$. The incoming $\phi(l)$ denotes a Fourier component of the collision background, while $\phi(p)$ and $h_{\mu\nu}(k)$ are physical final-state excitations.}
    \label{fig:two_body_scalar_graviton_tree}
\end{figure}

We first consider graviton emission accompanying the production of a scalar excitation from the collision background,
\begin{equation}
    \phi(l)\longrightarrow\phi(p)+h_{\mu\nu}(k).
    \label{eq:two_body_scalar_graviton_channel}
\end{equation}
Here \(\phi(l)\) denotes the classical collision background, while the outgoing \(\phi(p)\) is a physical on-shell excitation of the same scalar field and \(h_{\mu\nu}(k)\) is the emitted graviton. The corresponding tree-level contribution is shown in Fig.~\ref{fig:two_body_scalar_graviton_tree}.

For the two-body process, we take the collision axis along the background spatial momentum \(\boldsymbol{\ell}\) and write the four-momenta in the form
\begin{equation}
\begin{aligned}
    l^\mu&=(E,\boldsymbol{\ell})=(E,0,0,\ell),\\
    k^\mu&=(k,\boldsymbol{k})
    =k\bigl(1,\sigma\cos\theta,\sigma\sin\theta,\mu\bigr),\\
    p^\mu&=(E_p,\boldsymbol{p})=l^\mu-k^\mu\\
    &=\bigl(E_p,-k\sigma\cos\theta,-k\sigma\sin\theta,\ell-k\mu\bigr).
\end{aligned}
\label{eq:two_body_collision_frame_kinematics}
\end{equation}
where
\begin{equation}
    \mu\equiv\widehat{\boldsymbol{\ell}}\cdot\widehat{\boldsymbol{k}},
    \qquad
    \sigma\equiv\sqrt{1-\mu^2}.
    \label{eq:two_body_emission_direction_cosines}
\end{equation}
The metric convention is \(\eta_{\mu\nu}=\operatorname{diag}(+,-,-,-)\). The invariant squared four-momentum of the background mode and the on-shell conditions for the final-state particles are
\begin{equation}
    k_\mu k^\mu=0,\qquad p_\mu p^\mu=m_\phi^2.
    \label{eq:two_body_final_state_on_shell}
\end{equation}

For a physical outgoing graviton, we use the transverse-traceless polarization tensor \(\epsilon^{(\lambda)}_{\mu\nu}(k)\), which satisfies
\begin{equation}
    k^\mu\epsilon^{(\lambda)}_{\mu\nu}(k)=0,
    \qquad
    \eta^{\mu\nu}\epsilon^{(\lambda)}_{\mu\nu}(k)=0,
    \qquad
    \epsilon^{(\lambda)}_{\mu\nu}(k)=\epsilon^{(\lambda)}_{\nu\mu}(k).
    \label{eq:physical_graviton_tt_conditions}
\end{equation}
For an outgoing graviton the amplitude is contracted with \(\epsilon^{(\lambda)*}_{\mu\nu}(k)\). The sum over its two physical polarizations is
\begin{equation}
\begin{aligned}
    \Pi_{\mu\nu,\alpha\beta}(k)
    &\equiv
    \sum_{\lambda}
    \epsilon^{(\lambda)*}_{\mu\nu}(k)
    \epsilon^{(\lambda)}_{\alpha\beta}(k)
    \\
    &=\frac12\left(
        \hat\eta_{\mu\alpha}\hat\eta_{\nu\beta}
        +\hat\eta_{\mu\beta}\hat\eta_{\nu\alpha}
        -\hat\eta_{\mu\nu}\hat\eta_{\alpha\beta}
    \right),
\end{aligned}
\label{eq:physical_graviton_polarization_projector}
\end{equation}
where the reduced metric transverse to the graviton momentum is
\begin{equation}
    \hat\eta_{\mu\nu}
    \equiv
    \eta_{\mu\nu}
    -\frac{k_\mu\bar k_\nu+\bar k_\mu k_\nu}{k\cdot\bar k},
    \qquad
    \bar k^\mu\equiv(k,-\boldsymbol{k}).
    \label{eq:transverse_metric_projector}
\end{equation}
These relations are also collected in Appendix~\ref{app:graviton_polarization_sum}, while the corresponding graviton vertices are given in Appendix~\ref{app:graviton_vertices}.

Using the scalar-scalar-graviton vertex in Appendix~\ref{app:scalar_graviton_vertex}, we obtain the simplified polarized amplitude 
\begin{equation}
    \mathcal M_{\phi h}^{(\lambda)}
    =
    -\frac{2}{M_{\mathrm{pl}}}
    \epsilon^{(\lambda)*}_{\mu\nu}(k)l^\mu l^\nu.
    \label{eq:two_body_scalar_graviton_amplitude}
\end{equation}
Here and throughout the numerical evaluation, we use the reduced Planck mass \(M_{\mathrm{pl}}=2.44\times10^{15}~{\mathrm{TeV}}\).
Summing over the two physical graviton polarizations and using Eq.~\eqref{eq:physical_graviton_polarization_projector}, we obtain
\begin{equation}
    \overline{|\mathcal M_{\phi h}|^2}
    \equiv\sum_{\lambda}|\mathcal M_{\phi h}^{(\lambda)}|^2
    =\frac{2}{M_{\mathrm{pl}}^2}
    \left(\hat\eta_{\mu\nu} l^\mu l^\nu\right)^2
    =\frac{2}{M_{\mathrm{pl}}^2}\ell^4(1-\mu^2)^2,
    \label{eq:two_body_scalar_graviton_amplitude_squared}
\end{equation}
here $\hat\eta_{\mu\nu} l^\mu l^\nu=-\ell^2(1-\mu^2)$ denotes the transverse projection. Thus, Eq.~\eqref{eq:two_body_scalar_graviton_amplitude_squared} is the transverse-traceless projected squared amplitude for the process shown in Fig.~\ref{fig:two_body_scalar_graviton_tree}.

With the squared amplitude in hand,
\begin{equation}
    \left.\mathrm{Im}\,\widetilde\Gamma^{(2)}(E,\ell)\right|_{\phi h}
    =
    \frac12\int d\Pi_{\phi h}\overline{|\mathcal M_{\phi h}|^2}
    \Theta(\chi-m_\phi^2).
    \label{eq:two_body_imaginary_self_energy}
\end{equation}
where the physical two-body phase space
\begin{equation}
    d\Pi_{\phi h}=(2\pi)^4\delta^{(4)}(l-p-k)
    \frac{d^3\boldsymbol{p}}{(2\pi)^3\,2E_p}
    \frac{d^3\boldsymbol{k}}{(2\pi)^3\,2k}.
    \label{eq:two_body_lorentz_phase_space}
\end{equation}
The spatial delta function fixes \(\boldsymbol p=\boldsymbol\ell-\boldsymbol k\). The outgoing scalar energy is therefore \(E_p=\sqrt{m_\phi^2+\ell^2+k^2-2\ell k\mu}\). Writing the graviton momentum as \(d^3\boldsymbol k=k^2\,dk\,d\mu\,d\theta\), where \(\theta\) is the azimuthal angle about the collision axis and \(\mu\) is the polar direction cosine, both the squared amplitude and \(E_p\) are independent of \(\theta\), so the azimuthal integral gives
\begin{equation}
    d\Pi_{\phi h}=\frac{d^3\boldsymbol k}{4(2\pi)^2E_pk}\delta(E-E_p-k)=\frac{k\,dk\,d\mu}{8\pi E_p}\delta(E-E_p-k).
\label{eq:two_body_phase_space_after_azimuth}
\end{equation}
The remaining energy delta function fixes
\begin{equation}
\begin{aligned}
    &E_p=E-k=\sqrt{m_\phi^2+\ell^2+k^2-2\ell k\mu_0},\\
    &\mu_0=\frac{m_\phi^2-\chi+2Ek}{2\ell k},\\
    &\left|\frac{\partial(E-E_p-k)}{\partial\mu}\right|_{\mu_0}=\frac{\ell k}{E_p}.
\end{aligned}
\label{eq:two_body_energy_delta_root}
\end{equation}
The phase space therefore becomes 
\begin{equation}
    d\Pi_{\phi h}=\frac{dk\,d\mu}{8\pi}\frac{k}{E_p}\delta(E-E_p-k)=\frac{dk\,d\mu}{8\pi\ell}\delta(\mu-\mu_0)\Theta(1-|\mu_0|)\Theta(E-k).
\label{eq:two_body_phase_space_after_energy_delta}
\end{equation}
The factor \(\Theta(1-|\mu_0|)\) requires the root of the energy delta function to lie in the angular integration range \(-1\leq\mu\leq1\). The factor \(\Theta(E-k)\) enforces the positive-energy branch \(E_p=E-k>0\). It must be retained when solving the on-shell equation by squaring. The Jacobian above applies for \(\ell k>0\), with the endpoints understood by continuity. Performing the \(\mu\) integral, we obtain
\begin{equation}
    \left.\mathrm{Im}\,\widetilde\Gamma^{(2)}(E,\ell)\right|_{\phi h}=\frac{\ell^3}{8\pi M_{\mathrm{pl}}^2}\int_0^\infty dk\,(1-\mu_0^2)^2\Theta(\chi-m_\phi^2)\Theta(1-|\mu_0|)\Theta(E-k).
\label{eq:two_body_differential_imaginary_self_energy}
\end{equation}

Substituting Eq.~\eqref{eq:two_body_differential_imaginary_self_energy} into Eqs.~\eqref{eq:graviton_yield_per_collision_area} and \eqref{eq:thin_wall_collision_efficiency} then gives the number of gravitons produced in the two-body channel per unit area,
\begin{equation}
\begin{aligned}
    \frac{N_{\phi h}}{A}
    &=\frac{4v_\phi^2}{\pi^3M_{\mathrm{pl}}^2}
    \int_{\Lambda_{\min}}^{\Lambda_{\max}}dE
    \int_{\Lambda_{\min}}^{\Lambda_{\max}}d\ell\,
    \frac{\ell^3}{(E^2-\ell^2)^2}\\
    &\quad\times\int_0^\infty dk\,(1-\mu_0^2)^2
    \Theta(\chi-m_\phi^2)
    \Theta(1-|\mu_0|)\Theta(E-k).
\end{aligned}
\label{eq:two_body_graviton_yield_integral}
\end{equation}
The integrand with respect to \(dk\) gives the differential number spectrum.
The resulting differential number spectrum is per unit wall area, and its integral gives
\begin{equation}
    \frac{N_{\phi h}}{A}
    =
    \int_0^\infty dk\,
    \frac{d}{dk}\left(\frac{N_{\phi h}(k)}{A}\right).
    \label{eq:two_body_differential_graviton_number}
\end{equation}
To convert the number per unit area into a volume density, we approximate the colliding bubbles by two spheres of radius \(R_*\). This estimation applies when their size distribution is not parametrically broad. Assigning the flux in each longitudinal direction to the corresponding bubble volume gives
\begin{equation}
    \frac{A}{V}
    \simeq
    \frac{4\pi R_*^2}{2(4\pi R_*^3/3)}
    =\frac{3}{2R_*}.
    \label{eq:bubble_area_to_volume_ratio}
\end{equation}
For a broad bubble size distribution, \(3/(2R_*)\) should be replaced by the appropriate average area per unit volume. With the characteristic radius estimation, we obtain
\begin{equation}
    \frac{dn_{\mathrm{GW}}}{dk}
    =
    \frac{3}{2R_*}
    \frac{d}{dk}\left(\frac{N_{\phi h}(k)}{A}\right).
    \label{eq:two_body_graviton_number_density}
\end{equation}
Each graviton carries energy \(k\), so the GW energy density spectrum at production is
\begin{equation}
    \Omega_{\mathrm{GW}}^{*(\phi h)}(k)
    \equiv
    \frac{1}{\rho_{\mathrm{tot}}}
    \frac{d\rho_{\mathrm{GW}}^*}{d\ln k}
    =
    \frac{3k^2}{2R_*\rho_{\mathrm{tot}}}
    \frac{d}{dk}\left(\frac{N_{\phi h}(k)}{A}\right),
    \label{eq:two_body_gw_energy_spectrum_definition}
\end{equation}
where \(\rho_{\mathrm{tot}}\) is the total energy density at the transition, with its benchmark definition given below. Substituting Eq.~\eqref{eq:two_body_graviton_yield_integral} into Eq.~\eqref{eq:two_body_gw_energy_spectrum_definition}, one gets the production spectrum in a form that keeps the kinematic restrictions explicit through step functions,
\begin{equation}
\begin{aligned}
    \Omega_{\mathrm{GW}}^{*(\phi h)}(k)
    &={}
    \frac{6v_\phi^2k^2}
    {\pi^3R_*\rho_{\mathrm{tot}}M_{\mathrm{pl}}^2}
    \int_{\Lambda_{\min}}^{\Lambda_{\max}}d\ell\,\ell^3
    \int_{\Lambda_{\min}}^{\Lambda_{\max}}dE\,
    \frac{(1-\mu_0^2)^2}{(E^2-\ell^2)^2}
    \\
    &\quad\times
    \Theta(\chi-m_\phi^2)
    \Theta(1-|\mu_0|)\Theta(E-k).
\end{aligned}
\label{eq:two_body_gw_spectrum_step_function_form}
\end{equation}
The finite-bubble scale enters through the background window for \(E\) and \(\ell\).

For numerical evaluation and for making the analytical approximation below explicit, it is useful to replace the step functions in Eq.~\eqref{eq:two_body_gw_spectrum_step_function_form} by the corresponding kinematic integration region. For fixed \(E\) and \(k\), define
\begin{equation}
    \Delta(E,k)
    \equiv
    \sqrt{(E-k)^2-m_\phi^2}.
    \label{eq:two_body_scalar_recoil_momentum}
\end{equation}
The on-shell condition for the outgoing scalar gives
\begin{equation}
    \mu_0
    =
    \frac{\ell^2+k^2-\Delta^2(E,k)}
    {2\ell k}.
    \label{eq:two_body_emission_angle_root}
\end{equation}
The angular condition \(|\mu_0|\leq1\) is therefore equivalent to
\begin{equation}
    |k-\Delta(E,k)|
    \leq
    \ell
    \leq
    k+\Delta(E,k).
    \label{eq:two_body_longitudinal_kinematic_interval}
\end{equation}
After intersecting this interval with the background momentum window, the allowed range of \(\ell\) becomes
\begin{equation}
    \max\!\left\{
        \Lambda_{\min},
        |k-\Delta(E,k)|
    \right\}
    \leq
    \ell
    \leq
    k+\Delta(E,k),
    \label{eq:two_body_longitudinal_background_intersection}
\end{equation}
where the upper background bound is automatically satisfied because \(k+\Delta(E,k)<E\leq\Lambda_{\max}\).

The lower limit of the \(E\) integration follows from requiring the interval in Eq.~\eqref{eq:two_body_longitudinal_background_intersection} to be non negative. Defining
\begin{equation}
    \Delta_{\min}(k)
    \equiv
    \max\!\left\{0,\Lambda_{\min}-k\right\},
\end{equation}
the minimum allowed background energy is
\begin{equation}
    E_{\min}(k)
    =
    k+
    \sqrt{
        m_\phi^2+\Delta_{\min}^2(k)
    }.
    \label{eq:two_body_minimum_background_energy}
\end{equation}
The two-body spectrum can then be written as
\begin{equation}
\begin{aligned}
    \Omega_{\mathrm{GW}}^{*(\phi h)}(k)
    &={}
    \frac{6v_\phi^2k^2}
    {\pi^3R_*\rho_{\mathrm{tot}}M_{\mathrm{pl}}^2}
    \int_{E_{\min}(k)}^{\Lambda_{\max}}dE
    \int_{\max\{\Lambda_{\min},\,|k-\Delta(E,k)|\}}
         ^{k+\Delta(E,k)}
    d\ell\,\ell^3
    \frac{(1-\mu_0^2)^2}
    {(E^2-\ell^2)^2}.
\end{aligned}
\label{eq:two_body_gw_spectrum_kinematic_limits}
\end{equation}
This expression contributes when \(E_{\min}(k)<\Lambda_{\max}\). In particular, \(k<\Lambda_{\max}-m_\phi\) is a necessary kinematic condition.

We next relate the two-body spectrum to the present epoch and specify the phase transition parameters used for the numerical results. At the transition temperature \(T_*\), we take
\begin{equation}
\begin{gathered}
    \rho_{\mathrm{rad}}=\frac{\pi^2}{30}g_*T_*^4,
    \qquad
    \rho_{\mathrm{tot}}=(1+\alpha)\rho_{\mathrm{rad}},
    \qquad
    H_*=\sqrt{\frac{\rho_{\mathrm{tot}}}{3M_{\mathrm{pl}}^2}}.
\end{gathered}
\label{eq:transition_energy_density_and_hubble}
\end{equation}
Here, \(\rho_{\mathrm{rad}}\) and \(\rho_{\mathrm{tot}}\) are the radiation and total energy densities at the transition, respectively. The quantity \(g_*\) is the effective number of relativistic degrees of freedom, for which we use \(g_*=106.75\) in the numerical evaluation. The parameter \(\alpha\) is the transition strength, and \(H_*\) is the Hubble parameter at the transition. The inverse transition timescale \(\beta\) is specified by the dimensionless ratio \(\beta/H_*\). For the benchmark value \(\alpha=10\), the total energy densities is \(\rho_{\mathrm{tot}}=11\rho_{\mathrm{rad}}\). For the ultrarelativistic walls considered here, the characteristic collision scales used in the numerical calculation are
\begin{equation}
    R_*\simeq\frac{(8\pi)^{1/3}}{\beta},
    \qquad
    \Lambda_{\min}=R_*^{-1},
    \qquad
    \Lambda_{\max}=\gamma_{\mathrm{wall}}v_\phi.
    \label{eq:bubble_radius_and_collision_cutoffs}
\end{equation}
Here, the equalities for \(\Lambda_{\min}\) and \(\Lambda_{\max}\) specify the cutoff values adopted in the numerical integrations. The ultraviolet prescription follows from the order-of-magnitude estimate \(l_{{\mathrm{wall}},0}\sim v_\phi^{-1}\), which gives \(l_{\mathrm{wall}}\sim1/(\gamma_{\mathrm{wall}}v_\phi)\) and hence \(l_{\mathrm{wall}}^{-1}\sim\gamma_{\mathrm{wall}}v_\phi\). Thus Eqs.~\eqref{eq:transition_energy_density_and_hubble} and \eqref{eq:bubble_radius_and_collision_cutoffs} determine \(R_*\), \(\Lambda_{\min}\), and the adopted \(\Lambda_{\max}\) for each benchmark. The derivation of these scale relations and the validity conditions of the planar-wall approximation are given in Appendix~\ref{app:particle_production_formalism}. We assume that the released vacuum energy is rapidly converted into radiation after the transition, followed by radiation-dominated expansion.

After this conversion, we assume adiabatic expansion with no further entropy injection. At high temperature, we take \(g_{*\rho}\simeq g_{*s}\equiv g_*\). The standard radiation-era redshift relations used here are~\cite{Caprini:2019egz,Nakayama:2018pt,Barman:2023ymn,Barman:2023monomial}
\begin{equation}
    k_0=\frac{T_0}{T_*}\left(\frac{g_{*s,0}}{g_*}\right)^{1/3}k,\qquad h^2\Omega_{{\mathrm{GW}},0}(k_0)\simeq1.7\times10^{-5}\left(\frac{100}{g_*}\right)^{1/3}\Omega_{{\mathrm{GW}}}^{*}(k).
\label{eq:gw_redshift_to_present_epoch}
\end{equation}
Here, \(\Omega_{\mathrm{GW}}^{*}\) and \(\Omega_{{\mathrm{GW}},0}\) denote the GW spectrum at production and at the present epoch, respectively. We use the present photon temperature \(T_0=2.35\times10^{-16}~{\mathrm{TeV}}\), the present effective number of entropy degrees of freedom \(g_{*s,0}=3.931\), and the reduced Hubble constant \(h=0.674\). Defining the momentum redshift factor as \(r_s=(T_0/T_*)(g_{*s,0}/g_*)^{1/3}\), the first relation in Eq.~\eqref{eq:gw_redshift_to_present_epoch} becomes \(k_0=r_s k\). The corresponding present-day frequency is \(f_0=(2.418\times10^{26}~{\mathrm{Hz/TeV}})\,r_s k\). The numerical coefficient \(1.7\times10^{-5}\) in the second relation follows from the cosmological redshift of the GW energy density, which scales as \(a^{-4}\), and directly converts \(\Omega_{\mathrm{GW}}^{*}\) into \(h^2\Omega_{{\mathrm{GW}},0}\).

\begin{table*}[t]
\caption{Benchmark phase-transition parameters used in the numerical calculations. The parameter values varied in each comparison are specified in the corresponding figure captions.}
\label{tab:physical_cosmological_parameters}
\begin{ruledtabular}
\begin{tabular}{cccc}
\(\alpha\) & \(\beta/H_*\) & \(T_*\) & \(m_\phi\) \\
\hline
\(10\) & \(100\) & \(0.1v_\phi\) & \(0.5v_\phi\) \\
\end{tabular}
\end{ruledtabular}
\end{table*}

The wall Lorentz factors are chosen to remain below the order-of-magnitude runaway bound
\begin{equation}
    \gamma_{\mathrm{wall}}^{\max}\sim
    \frac{H_*}{\beta}\frac{M_{\mathrm P}}{v_\phi}
    =
    \frac{M_{\mathrm P}}{(\beta/H_*)v_\phi},
    \label{eq:runaway_wall_lorentz_bound}
\end{equation}
where \(M_{\mathrm P}\simeq1.22\times10^{19}~{\rm GeV}\) is the conventional Planck mass, distinct from the reduced Planck mass \(M_{\mathrm{pl}}\) used in the graviton couplings and amplitudes. For \(v_\phi=10^{10}~{\rm GeV}\), Eq.~\eqref{eq:runaway_wall_lorentz_bound} gives \(\gamma_{\mathrm{wall}}^{\max}\simeq1.22\times10^7\). Thus, at fixed \(v_\phi\), all values considered up to \(\gamma_{\mathrm{wall}}=10^7\) remain below this estimate. Likewise, for fixed \(\gamma_{\mathrm{wall}}=10^7\), the values \(v_\phi=10^8,10^9,10^{10}~{\rm GeV}\) all satisfy the same bound. The most restrictive point is \(v_\phi=10^{10}~{\rm GeV}\), where \(\gamma_{\mathrm{wall}}^{\max}\simeq1.22\times10^7\).

\begin{figure}[t]
    \centering
    \includegraphics[width=\columnwidth]{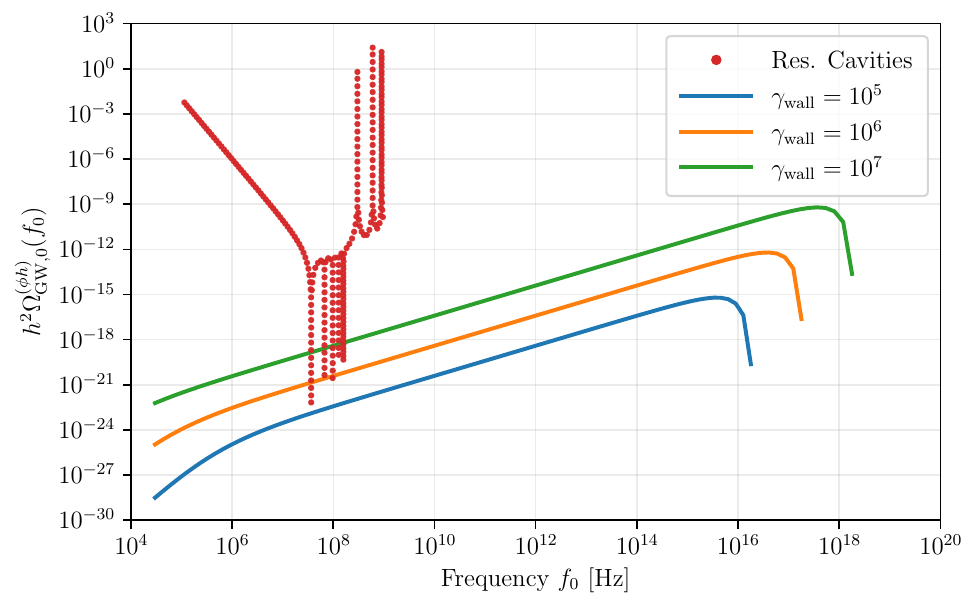}
    \caption{Present-day GW spectrum \(h^2\Omega_{{\mathrm{GW}},0}^{(\phi h)}\) for the two-body channel \(\phi(l)\to\phi(p)+h_{\mu\nu}(k)\) at fixed \(v_\phi=10^{10}~{\rm GeV}\), with \(T_*=10^9~{\rm GeV}\) and \(m_\phi=5\times10^9~{\rm GeV}\). The curves correspond to \(\gamma_{\mathrm{wall}}=10^5,10^6,10^7\).}
    \label{fig:two_body_gamma_wall_dependence}
\end{figure}

Using the common benchmark relations in Table~\ref{tab:physical_cosmological_parameters}, we consider two complementary parameter variations. First, we fix \(v_\phi\) and vary the wall Lorentz factor over the values shown in Fig.~\ref{fig:two_body_gamma_wall_dependence}. In this comparison, \(\Lambda_{\min}\) and the cosmological redshift are fixed, while the ultraviolet scale changes as \(\Lambda_{\max}=\gamma_{\mathrm{wall}}v_\phi\). For reference, the spectrum figures below also overlay projected sensitivities of resonant-cavity GW searches~\cite{Aggarwal:2020olq,Herman:2023hfgw}. These sensitivity markers are included only for qualitative comparison and should not be interpreted as detector forecasts.

Figure~\ref{fig:two_body_gamma_wall_dependence} illustrates the dependence of the two-body GW spectrum on the wall Lorentz factor. Increasing \(\gamma_{\mathrm{wall}}\) raises the ultraviolet scale \(\Lambda_{\max}=\gamma_{\mathrm{wall}}v_\phi\) linearly, broadens the range of collision-background modes entering the \(E\) and \(\ell\) integrations, and extends the spectrum to higher frequencies, while \(T_*\), \(m_\phi\), \(H_*\), and \(\Lambda_{\min}\) remain unchanged. Near the upper endpoint, the spectrum drops rapidly for kinematic reasons. As \(k\) approaches \(\Lambda_{\max}-m_\phi\), the outgoing scalar must still retain at least its rest energy, so the allowed region of the \((E,\ell)\) integration in Eq.~\eqref{eq:two_body_gw_spectrum_kinematic_limits} continuously shrinks and vanishes at the endpoint.

\begin{figure}[t]
    \centering
    \includegraphics[width=\columnwidth]{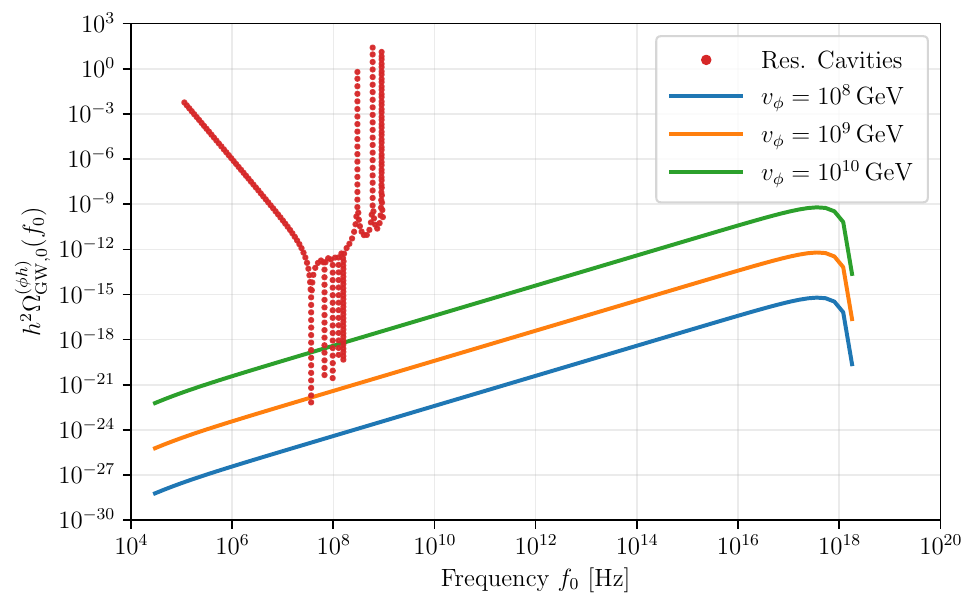}
    \caption{Present-day GW spectrum for the two-body channel at fixed \(\gamma_{\mathrm{wall}}=10^7\), varying \(v_\phi=10^{8},10^{9},10^{10}~{\rm GeV}\). For each curve, the correlated benchmark relations \(T_*=0.1v_\phi\) and \(m_\phi=0.5v_\phi\) are imposed.}
    \label{fig:two_body_transition_scale_dependence}
\end{figure}

For the complementary parameter variation, we instead fix \(\gamma_{\mathrm{wall}}\) and vary the transition scale \(v_\phi\) over the values shown in Fig.~\ref{fig:two_body_transition_scale_dependence}. The temperature and scalar mass follow the benchmark relations in Table~\ref{tab:physical_cosmological_parameters}.

At fixed \(\gamma_{\mathrm{wall}}\), changing \(v_\phi\) modifies several physical scales simultaneously. Since \(T_*\propto v_\phi\), one has \(H_*\propto v_\phi^2\), \(\beta\propto v_\phi^2\), and \(\Lambda_{\min}\propto v_\phi^2\), whereas \(\Lambda_{\max}=\gamma_{\mathrm{wall}}v_\phi\propto v_\phi\). The ratio \(m_\phi/\Lambda_{\max}=1/(2\gamma_{\mathrm{wall}})\) is therefore fixed throughout this variation. Moreover, the redshifted ultraviolet scale is approximately independent of \(v_\phi\), because \(T_*=0.1v_\phi\) implies \(f_{0,\mathrm{UV}}\propto\Lambda_{\max}/T_*\propto\gamma_{\mathrm{wall}}\). By contrast, the overall normalization and the lower collision scale change with \(v_\phi\). In the linear regime described analytically below, the leading two-body normalization scales as \(v_\phi^4/T_*\propto v_\phi^3\) at fixed \(\gamma_{\mathrm{wall}}\). Taken together, Figs.~\ref{fig:two_body_gamma_wall_dependence} and~\ref{fig:two_body_transition_scale_dependence} separate the effects of the wall boost from those of the overall phase-transition scale while preserving the Higgs-like relations among \(v_\phi\), \(T_*\), and \(m_\phi\).

The approximately linear rise of the two-body spectrum can be understood analytically from the exact two-body kinematics. In the intermediate momentum region, the leading contribution to the asymptotic expansion is
\begin{equation}
    \Omega_{\mathrm{GW}}^{*(\phi h)}(k)
    \simeq
    \frac{2v_\phi^2\Lambda_{\max}^2}{\pi^3R_*\rho_{\mathrm{tot}}M_{\mathrm{pl}}^2}\,k.
\label{eq:two_body_linear_momentum_scaling}
\end{equation}
Thus the two-body GW spectrum rises linearly with the emitted graviton momentum. The derivation and the subleading terms are given in Appendix~\ref{app:two_body_asymptotic}. The leading-order approximation applies in the parametric window
\begin{equation}
    \frac{m_\phi^2}{\Lambda_{\max}}\ll k\ll\Lambda_{\max}.
\label{eq:two_body_linear_regime_momentum_window}
\end{equation}
The lower condition ensures that the scalar-mass-dependent corrections remain subleading, while the upper condition keeps the emitted momentum well below the ultraviolet background scale.

To express this linear behavior in terms of the present-day frequency, we use Eq.~\eqref{eq:gw_redshift_to_present_epoch}. With \(\Lambda_{\max}=\gamma_{\mathrm{wall}}v_\phi\), \(R_*\simeq(8\pi)^{1/3}/\beta\), \(\beta=(\beta/H_*)H_*\), and \(\rho_{\mathrm{tot}}=(1+\alpha)\pi^2g_*T_*^4/30\), Eq.~\eqref{eq:two_body_linear_momentum_scaling} becomes
\begin{equation}
\begin{aligned}
    h^2\Omega_{{\mathrm{GW}},0}^{(\phi h)}(f_0)
    \simeq
    &3.92\times10^{-17}
    \left(\frac{1+\alpha}{11}\right)^{-1/2}
    \left(\frac{\beta/H_*}{100}\right)
    \left(\frac{\gamma_{\mathrm{wall}}}{10^7}\right)^2\\
    &\left(\frac{v_\phi}{10^{10}\mathrm{~GeV}}\right)^4
    \left(\frac{T_*}{10^{9}\mathrm{~GeV}}\right)^{-1}
    \left(\frac{f_0}{10^{10}\mathrm{~Hz}}\right).
\end{aligned}
\label{eq:two_body_present_day_linear_scaling}
\end{equation}

The momentum window in Eq.~\eqref{eq:two_body_linear_regime_momentum_window} maps directly onto the present-day frequency range
\begin{equation}
    f_{0,m}\ll f_0\ll f_{0,\mathrm{UV}},
    \qquad
    f_{0,m}\equiv
    \left(\frac{m_\phi}{\Lambda_{\max}}\right)^2 f_{0,\mathrm{UV}},
\label{eq:two_body_linear_regime_frequency_window}
\end{equation}
where the redshifted ultraviolet scale is
\begin{equation}
\begin{aligned}
    f_{0,\mathrm{UV}}
    &\equiv(2.418\times10^{26}~\mathrm{Hz/TeV})\,r_s\Lambda_{\max}\\
    &\simeq1.89\times10^{18}
    \left(\frac{\gamma_{\mathrm{wall}}}{10^7}\right)
    \left(\frac{v_\phi}{10^{10}\mathrm{~GeV}}\right)
    \left(\frac{T_*}{10^{9}\mathrm{~GeV}}\right)^{-1}
    ~\mathrm{Hz}.
\end{aligned}
\label{eq:two_body_redshifted_uv_frequency}
\end{equation}
For the benchmark relations \(T_*=0.1v_\phi\) and \(m_\phi=0.5v_\phi\), one has \(f_{0,m}=f_{0,\mathrm{UV}}/(4\gamma_{\mathrm{wall}}^2)\). For the fixed-\(v_\phi\) comparison in Fig.~\ref{fig:two_body_gamma_wall_dependence}, the characteristic linear regions are therefore
\begin{equation}
\begin{aligned}
    \gamma_{\mathrm{wall}}=10^5:&\qquad
    4.73\times10^5~\mathrm{Hz}\ll f_0\ll1.89\times10^{16}~\mathrm{Hz},\\
    \gamma_{\mathrm{wall}}=10^6:&\qquad
    4.73\times10^4~\mathrm{Hz}\ll f_0\ll1.89\times10^{17}~\mathrm{Hz},\\
    \gamma_{\mathrm{wall}}=10^7:&\qquad
    4.73\times10^3~\mathrm{Hz}\ll f_0\ll1.89\times10^{18}~\mathrm{Hz}.
\end{aligned}
\label{eq:two_body_benchmark_linear_frequency_ranges}
\end{equation}
For the fixed \(\gamma_{\mathrm{wall}}=10^7\) comparison in Fig.~\ref{fig:two_body_transition_scale_dependence}, the same characteristic frequency window applies to all three values of \(v_\phi\), because the benchmark relation \(T_*=0.1v_\phi\) is fixed. The inequalities in Eqs.~\eqref{eq:two_body_linear_regime_momentum_window} and \eqref{eq:two_body_linear_regime_frequency_window} are parametric conditions, so the leading linear approximation is expected to be accurate well inside, rather than directly at, these characteristic boundaries. The spectra shown in Figs.~\ref{fig:two_body_gamma_wall_dependence} and~\ref{fig:two_body_transition_scale_dependence} are evaluated from the full numerical expression.

\section{Three-body production with graviton emission}\label{sec:three_body_emission}

\begin{figure}[t]
\centering
\includegraphics[width=0.95\columnwidth]{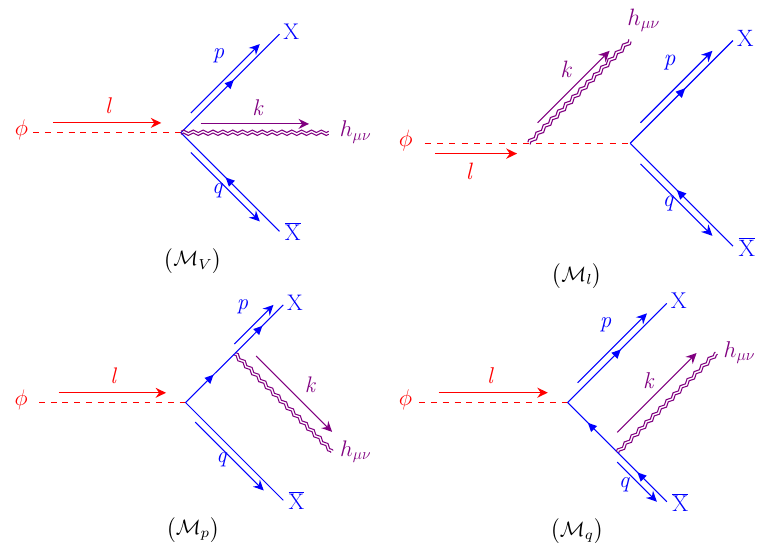}
\caption{Four tree-level contributions to the three-body process $\phi(l)\to X(p)+\bar X(q)+h_{\mu\nu}(k)$, with $X\bar X$ representing either $\varphi\varphi^\ast$ or $\psi\bar\psi$. Notice that the first diagram's contribution ($\mathcal M_{V}$) vanishes because the physical graviton polarization tensor is traceless.}
\label{fig:three_body_particle_graviton_tree_topologies}
\end{figure}

The collision background couples to the scalar and fermion fields through
\begin{equation}
    \mathcal{L}_{\mathrm{int}}
    =
    \begin{cases}
    -\lambda_\varphi \varphi^\ast\varphi\phi,
        & \text{scalar channel},\\
    -\lambda_\psi \bar{\psi}\psi\phi,
        & \text{fermion channel},
    \end{cases}
    \label{eq:scalar_fermion_background_interactions}
\end{equation}
where \(\lambda_\varphi\) has mass dimension one and \(\lambda_\psi\) is dimensionless. These interactions allow the collision background to produce scalar or fermion particle--antiparticle pairs, which might be the possible new dark matter production mechanism~\cite{Cheng:2026npt}. Including the universal gravitational interaction opens the corresponding three-body graviton-emission channels
\begin{equation}
    \phi(l)\to \varphi(p)+\varphi^\ast(q)+h_{\mu\nu}(k),
    \qquad
    \phi(l)\to \psi(p)+\bar\psi(q)+h_{\mu\nu}(k).
    \label{eq:scalar_fermion_three_body_channels}
\end{equation}
The two channels can be represented collectively as
\begin{equation}
    \phi(l)\to X(p)+\bar X(q)+h_{\mu\nu}(k).
    \label{eq:generic_three_body_particle_graviton_channel}
\end{equation}
with \(X(p)\bar X(q)\) denoting either \(\varphi(p)\varphi^\ast(q)\) or \(\psi(p)\bar\psi(q)\). The corresponding four possible tree-level contributions are shown in Fig.~\ref{fig:three_body_particle_graviton_tree_topologies}.

We adopt the same approximations and collision frame as in the two-body calculation of Sec.~\ref{sec:two_body_emission}. The background momentum is aligned with the \(z\) axis, and the four-momenta are written as
\begin{equation}
\begin{aligned}
    l^\mu&=(E,\boldsymbol{\ell})=(E,0,0,\ell),\\
    k^\mu&=(k,\boldsymbol{k})
    =k\bigl(1,\sigma_k\cos\theta_k,\sigma_k\sin\theta_k,\mu_k\bigr),
    \qquad
    \mu_k\equiv\widehat{\boldsymbol\ell}\cdot\widehat{\boldsymbol k},
    \qquad
    \sigma_k\equiv\sqrt{1-\mu_k^2},
    \\
    p^\mu&=(E_p,\boldsymbol{p})
    =\bigl(E_p,p\sigma_p\cos\theta_p,
    p\sigma_p\sin\theta_p,p\mu_p\bigr),
    \qquad
    \mu_p\equiv\widehat{\boldsymbol\ell}\cdot\widehat{\boldsymbol p},
    \qquad
    \sigma_p\equiv\sqrt{1-\mu_p^2},
    \\
    q^\mu&=(E_q,\boldsymbol{q})=l^\mu-p^\mu-k^\mu\\
    &=\bigl(E-k-E_p,
    -k\sigma_k\cos\theta_k-p\sigma_p\cos\theta_p,-k\sigma_k\sin\theta_k-p\sigma_p\sin\theta_p,
    \ell-k\mu_k-p\mu_p\bigr).
\end{aligned}
\label{eq:three_body_collision_frame_kinematics}
\end{equation}
Here \(m_X\) denotes the mass of the scalar or fermion in the corresponding final state. The momenta satisfy
\begin{equation}
    p_\mu p^\mu=q_\mu q^\mu=m_X^2,
    \qquad
    k_\mu k^\mu=0.
\label{eq:three_body_final_state_on_shell}
\end{equation}
Notice that the contact contribution vanishes because the physical graviton polarization tensor is traceless,
\begin{equation}
    \eta^{\mu\nu}\epsilon^{(\lambda)*}_{\mu\nu}(k)=0,
    \qquad
    \mathcal M_{V}^{(\lambda)}=0.
\label{eq:three_body_traceless_contact_vanishing}
\end{equation}
\noindent\textbf{Scalar channel.} The diagram-by-diagram derivation is given in Appendix~\ref{app:scalar_three_body_calculation}. In the main text we therefore retain only the three nonvanishing contributions in Fig.~\ref{fig:three_body_particle_graviton_tree_topologies},
\begin{equation}
\begin{aligned}
    \mathcal M_{\varphi\varphi^\ast h}^{(\lambda)}
    ={}&\mathcal M_{\varphi\varphi^\ast h,l}^{(\lambda)}
    +\mathcal M_{\varphi\varphi^\ast h,p}^{(\lambda)}
    +\mathcal M_{\varphi\varphi^\ast h,q}^{(\lambda)}\\
    ={}&\frac{\lambda_\varphi}{M_{\mathrm{pl}}}\,
    \epsilon^{(\lambda)*}_{\mu\nu}(k)
    \left(
    -\frac{l^\mu l^\nu}{D_l}
    +\frac{p^\mu p^\nu}{D_p}
    +\frac{q^\mu q^\nu}{D_q}
    \right).
\end{aligned}
\label{eq:scalar_pair_graviton_amplitude}
\end{equation}
The propagator denominators are
\begin{equation}
\begin{aligned}
    D_l
    &\equiv l\cdot k-\frac{\chi-m_\phi^2}{2}
    =k(E-\ell\mu_k)-\frac{\chi-m_\phi^2}{2},\\
    D_p
    &\equiv p\cdot k
    =k\!\left(E_p-p\,\widehat{\boldsymbol p}\!\cdot\!\widehat{\boldsymbol k}\right),\\
    D_q
    &\equiv q\cdot k
    =l\cdot k-p\cdot k
    =k\!\left(E-E_p-\ell\mu_k
    +p\,\widehat{\boldsymbol p}\!\cdot\!\widehat{\boldsymbol k}\right).
\end{aligned}
\label{eq:scalar_three_body_propagator_denominators}
\end{equation}
Here \(m_\phi\) is the mass parameter in the internal background-scalar propagator. To write the polarization-summed result compactly, define
\begin{equation}
    \mathcal T(a,b)
    \equiv
    \hat\eta_{\mu\nu}a^\mu b^\nu
    =-\boldsymbol a\cdot\boldsymbol b
    +(\widehat{\boldsymbol k}\cdot\boldsymbol a)
     (\widehat{\boldsymbol k}\cdot\boldsymbol b),
    \qquad
    \mathcal T(k,a)=0,
    \label{eq:three_body_transverse_contraction}
\end{equation}
and
\begin{equation}
\begin{aligned}
    \mathcal T_1
    &\equiv \mathcal T(l,l)
    =-\ell^2(1-\mu_k^2)
    =-\ell^2\sigma_k^2,\\
    \mathcal T_2
    &\equiv \mathcal T(p,p)
    =-p^2\!\left[1-
    \left(\widehat{\boldsymbol p}\!\cdot\!\widehat{\boldsymbol k}\right)^2\right],\\
    \mathcal T_3
    &\equiv \mathcal T(l,p)
    =\ell p\!\left[
    \mu_k\left(\widehat{\boldsymbol p}\!\cdot\!\widehat{\boldsymbol k}\right)-\mu_p
    \right].
\end{aligned}
\label{eq:three_body_transverse_invariants}
\end{equation}
Since \(q=l-p-k\) and \(\mathcal T(k,a)=0\),
\begin{equation}
    \mathcal T(q,q)=\mathcal T_1+\mathcal T_2-2\mathcal T_3,
    \qquad
    \mathcal T(l,q)=\mathcal T_1-\mathcal T_3,
    \qquad
    \mathcal T(p,q)=\mathcal T_3-\mathcal T_2.
\label{eq:three_body_q_transverse_relations}
\end{equation}
The scalar squared amplitude, summed over the two physical graviton polarizations, is
\begin{equation}
\begin{aligned}
\overline{|\mathcal M_{\varphi\varphi^\ast h}|^2}
=\left(\frac{\lambda_\varphi}{M_{\mathrm{pl}}}\right)^2
&\Bigg[
\frac{\mathcal T_1^2}{2D_l^2}
+\frac{\mathcal T_2^2}{2D_p^2}
+\frac{(\mathcal T_1+\mathcal T_2-2\mathcal T_3)^2}{2D_q^2}
-\frac{2\mathcal T_3^2-\mathcal T_1\mathcal T_2}{D_lD_p}\\
&-\frac{\mathcal T_1^2-2\mathcal T_1\mathcal T_3+2\mathcal T_3^2-\mathcal T_1\mathcal T_2}{D_lD_q}
+\frac{2\mathcal T_3^2-2\mathcal T_2\mathcal T_3+\mathcal T_2^2-\mathcal T_1\mathcal T_2}{D_pD_q}
\Bigg].
\end{aligned}
\label{eq:scalar_pair_graviton_amplitude_squared}
\end{equation}

\noindent\textbf{Fermion channel.} The corresponding diagram-by-diagram reduction is given in Appendix~\ref{app:fermion_three_body_calculation}. After combining the four diagrams, the amplitude can be written in the reduced form
\begin{equation}
\begin{aligned}
\mathcal M_{\psi\bar\psi h}^{(\lambda)}
={}&\mathcal M_{\psi\bar\psi h,l}^{(\lambda)}
+\mathcal M_{\psi\bar\psi h,p}^{(\lambda)}
+\mathcal M_{\psi\bar\psi h,q}^{(\lambda)}\\
={}&\frac{\lambda_\psi}{M_{\mathrm{pl}}}\,
\epsilon^{(\lambda)*}_{\mu\nu}(k)
\Bigg\{
\left(-\frac{l^\mu l^\nu}{D_l}
+\frac{p^\mu p^\nu}{D_p}
+\frac{q^\mu q^\nu}{D_q}\right)\bar u(p)v(q)\\
&\hspace{4.5em}
+\left(\frac{p^\mu}{2D_p}-\frac{q^\mu}{2D_q}\right)
\bar u(p)\gamma^\nu\slashed k\,v(q)
\Bigg\}.
\end{aligned}
\label{eq:fermion_pair_graviton_amplitude}
\end{equation}
The additional invariant entering the fermionic interference terms is
\begin{equation}
    \Sigma_\psi
    \equiv
    2(p\cdot q-m_\psi^2)+l\cdot k
    =\chi-4m_\psi^2-l\cdot k.
\label{eq:fermion_interference_invariant}
\end{equation}
Summing over the final-state fermion spins and the two graviton polarizations gives
\begin{equation}
\label{eq:fermion_pair_graviton_amplitude_squared}
\begin{aligned}
\overline{|\mathcal M_{\psi\bar\psi h}|^2}
=\frac{\lambda_\psi^2}{M_{\mathrm{pl}}^2}
\Bigg\{
&\frac{(\chi-4m_\psi^2-2l\cdot k)\mathcal T_1^2}{D_l^2}
+\frac{\mathcal T_2}{D_p^2}
\left[(\chi-4m_\psi^2)\mathcal T_2
-2D_p(\mathcal T_3+D_q)\right]\\
&+\frac{\mathcal T_1+\mathcal T_2-2\mathcal T_3}{D_q^2}
\left[(\chi-4m_\psi^2)(\mathcal T_1+\mathcal T_2-2\mathcal T_3)
-2D_q(\mathcal T_1-\mathcal T_3+D_p)\right]\\
&+\frac{1}{D_lD_p}
\left[-4\Sigma_\psi
\left(\mathcal T_3^2-\frac12\mathcal T_1\mathcal T_2\right)
+2D_p\mathcal T_1\mathcal T_3\right]\\
&+\frac{1}{D_lD_q}
\Bigg[-4\Sigma_\psi
\left((\mathcal T_1-\mathcal T_3)^2
-\frac12\mathcal T_1(\mathcal T_1+\mathcal T_2-2\mathcal T_3)\right)
+2D_q\mathcal T_1(\mathcal T_1-\mathcal T_3)\Bigg]\\
&+\frac{1}{D_pD_q}
\Bigg[4\Sigma_\psi
\left((\mathcal T_3-\mathcal T_2)^2
-\frac12\mathcal T_2(\mathcal T_1+\mathcal T_2-2\mathcal T_3)\right)\\
&\hspace{1.6cm}
-2(\mathcal T_3-\mathcal T_2)
\left(\mathcal T_2D_q
+(\mathcal T_1+\mathcal T_2-2\mathcal T_3)D_p
-2D_pD_q\right)\Bigg]
\Bigg\}.
\end{aligned}
\end{equation}

The squared amplitudes depend on the azimuthal angles \(\theta_k\) and \(\theta_p\) only through their relative angle,
\begin{equation}
    \Delta\theta\equiv\theta_p-\theta_k,
    \qquad
    \zeta\equiv
    \widehat{\boldsymbol p}\cdot\widehat{\boldsymbol k}
    =\mu_p\mu_k+\sigma_p\sigma_k\cos\Delta\theta.
    \label{eq:three_body_relative_azimuth}
\end{equation}
Thus the azimuthal dependence is carried by \(\cos\Delta\theta\), or equivalently by \(\zeta\).

The three-body phase-space reduction follows the same treatment as in the two-body channel, with an additional integration over the second matter momentum. For both the scalar and fermion processes, the Lorentz-invariant phase space is
\begin{equation}
    d\Pi_{X\bar X h}
    =
    (2\pi)^4\delta^{(4)}(l-p-q-k)
    \frac{d^3\boldsymbol p}{(2\pi)^3\,2E_p}
    \frac{d^3\boldsymbol q}{(2\pi)^3\,2E_q}
    \frac{d^3\boldsymbol k}{(2\pi)^3\,2k}.
    \label{eq:three_body_lorentz_phase_space}
\end{equation}
The corresponding contribution to the imaginary part of the background two-point function is fixed by the optical theorem,
\begin{equation}
    \left.\mathrm{Im}\,\widetilde\Gamma^{(2)}(E,\ell)\right|_{X\bar Xh}
    =
    \frac12\int d\Pi_{X\bar X h}\,
    \overline{|\mathcal M_{X\bar Xh}|^2},
    \label{eq:three_body_imaginary_self_energy}
\end{equation}
where the overline denotes the scalar polarization sum or the fermion spin--polarization sum defined above. The pair invariant mass entering the threshold condition is
\begin{equation}
    s_{X\bar X}\equiv(p+q)_\mu(p+q)^\mu=(l-k)_\mu(l-k)^\mu=\chi-2k(E-\ell\mu_k),\qquad s_{X\bar X}\geq4m_X^2.
\label{eq:three_body_pair_invariant_mass_threshold}
\end{equation}
The spatial delta function fixes \(\boldsymbol q=\boldsymbol\ell-\boldsymbol p-\boldsymbol k\), while axial symmetry removes the common azimuthal angle. The remaining energy delta function fixes the relative angle \(\Delta\theta\) defined in Eq.~\eqref{eq:three_body_relative_azimuth}. On the energy-conserving solution,
\begin{equation}
\begin{aligned}
    \zeta_0
    &=
    \frac{
    (E-E_p-k)^2-m_X^2-\ell^2-p^2-k^2
    +2\ell p\mu_p+2\ell k\mu_k
    }{2pk},
    \\
    u_0
    &\equiv \cos\Delta\theta_0
    =
    \frac{\zeta_0-\mu_p\mu_k}{\sigma_p\sigma_k}.
\end{aligned}
\label{eq:three_body_energy_conserving_angular_roots}
\end{equation}
Carrying out these reductions gives
\begin{equation}
\begin{aligned}
\left.\mathrm{Im}\,\widetilde\Gamma^{(2)}(E,\ell)\right|_{X\bar Xh}
={}&\frac{1}{8(2\pi)^4}
\int_0^\infty dk
\int_{-1}^{1}d\mu_k\,
\Theta(s_{X\bar X}-4m_X^2)\\
&\times\int_{-1}^{1}d\mu_p
\int_{m_X}^{E-k-m_X}dE_p\\
&\times\frac{
\left.\overline{|\mathcal M_{X\bar Xh}|^2}
\right|_{\cos\Delta\theta=u_0}}
{\sigma_k\sigma_p\sqrt{1-u_0^2}}
\Theta(1-u_0^2).
\end{aligned}
\label{eq:three_body_differential_imaginary_self_energy}
\end{equation}
The intermediate Jacobian steps are given in Appendix~\ref{app:three_body_phase_space}. 
In Eq.~\eqref{eq:three_body_differential_imaginary_self_energy}, the squared amplitudes are evaluated on the energy-conserving solution by setting \(\widehat{\boldsymbol p}\!\cdot\!\widehat{\boldsymbol k}=\zeta_0\), equivalently \(\cos\Delta\theta=u_0\), in Eqs.~\eqref{eq:scalar_three_body_propagator_denominators} and \eqref{eq:three_body_transverse_invariants}. The \(E_p\) integral vanishes when \(E-k<2m_X\). Substituting \(\overline{|\mathcal M_{\varphi\varphi^\ast h}|^2}\) from Eq.~\eqref{eq:scalar_pair_graviton_amplitude_squared} together with the scalar mass gives the scalar contribution, while substituting \(\overline{|\mathcal M_{\psi\bar\psi h}|^2}\) from Eq.~\eqref{eq:fermion_pair_graviton_amplitude_squared} together with the fermion mass gives the fermion contribution.

Using the collision factor \(f_{\mathrm{coll}}(E,\ell)\) in Eq.~\eqref{eq:thin_wall_collision_efficiency}, the differential graviton number spectrum per unit area is
\begin{equation}
\begin{aligned}
\frac{dN_{X\bar Xh}}{A\,dk}
={}&\frac{v_\phi^2}{4\pi^6}
\int_{\Lambda_{\min}}^{\Lambda_{\max}} dE
\int_{\Lambda_{\min}}^{\Lambda_{\max}} d\ell\,
\frac{\Theta(\chi)}{\chi^2}\\
&\times\int_{-1}^{1}d\mu_k\,
\Theta(s_{X\bar X}-4m_X^2)
\int_{-1}^{1}d\mu_p\\
&\times\int_{m_X}^{E-k-m_X}dE_p\,
\frac{\left.
\overline{|\mathcal M_{X\bar Xh}|^2}
\right|_{\cos\Delta\theta=u_0}}
{\sigma_k\sigma_p\sqrt{1-u_0^2}}
\Theta(1-u_0^2),\\
&\hspace{4.3em}0<k<\Lambda_{\max}-2m_X.
\end{aligned}
\label{eq:three_body_differential_graviton_yield}
\end{equation}
The pair-production threshold and angular constraints are contained explicitly in the step functions, while \(\Theta(\chi)\) restricts the collision background to timelike modes with \(\chi>0\).

Energy conservation requires \(E\geq k+2m_X\), which together with \(E\leq\Lambda_{\max}\) gives the upper bound \(k<\Lambda_{\max}-2m_X\), with \(m_X\) understood as the mass of the particle in the corresponding scalar or fermion final state. The background window and the phase-space constraints determine the support of the spectrum. Integrating the differential spectrum gives
\begin{equation}
    \frac{N_{X\bar Xh}}{A}
    =
    \int_0^\infty dk\,
    \frac{dN_{X\bar Xh}}{A\,dk}.
    \label{eq:three_body_total_graviton_yield}
\end{equation}

Using the same area-to-volume conversion as in the two-body channel, the GW spectrum at production is
\begin{equation}
    \Omega_{\mathrm{GW}}^{*(X\bar Xh)}(k)=\frac{3k^2}{2R_*\rho_{\mathrm{tot}}}\frac{dN_{X\bar Xh}}{A\,dk},\qquad 0<k<\Lambda_{\max}-2m_X.
\label{eq:three_body_gw_energy_spectrum}
\end{equation}

\begin{figure}[t]
    \centering
    \includegraphics[width=\columnwidth]{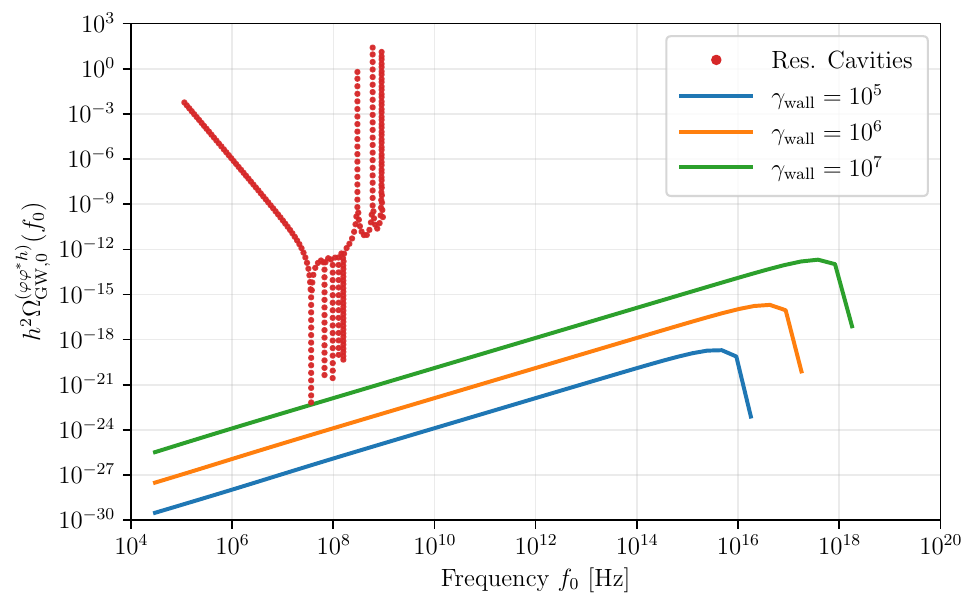}
    \caption{Present-day GW spectrum \(h^2\Omega_{{\mathrm{GW}},0}^{(\varphi\varphi^\ast h)}\) for the scalar three-body channel at fixed \(v_\phi=10^{10}\,\mathrm{GeV}\), with \(m_\varphi=5\times10^9\,\mathrm{GeV}\) and \(\lambda_\varphi=2.5\times10^9\,\mathrm{GeV}\). The curves correspond to \(\gamma_{\mathrm{wall}}=10^5,10^6,10^7\).}
    \label{fig:scalar_three_body_gamma_wall_dependence}
\end{figure}

\begin{figure}[t]
    \centering
    \includegraphics[width=\columnwidth]{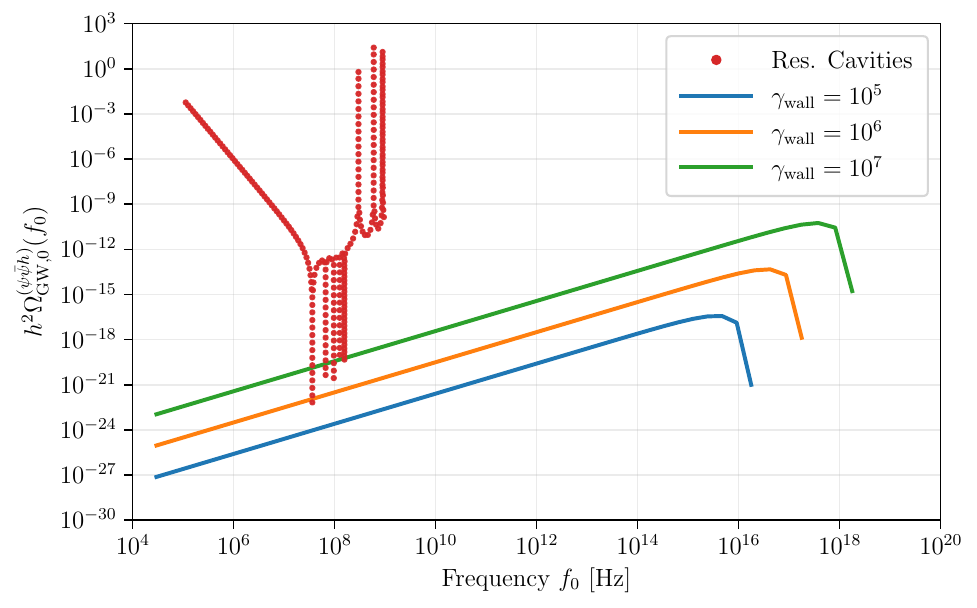}
    \caption{Present-day GW spectrum \(h^2\Omega_{{\mathrm{GW}},0}^{(\psi\bar\psi h)}\) for the fermion three-body channel at fixed \(v_\phi=10^{10}\,\mathrm{GeV}\), with \(m_\psi=5\times10^9\,\mathrm{GeV}\) and \(\lambda_\psi=0.5\). The curves correspond to \(\gamma_{\mathrm{wall}}=10^5,10^6,10^7\).}
    \label{fig:fermion_three_body_gamma_wall_dependence}
\end{figure}

The scalar and fermion channels differ through their squared amplitudes and the corresponding particle masses entering the phase space, leading to the production spectra \(\Omega_{\mathrm{GW}}^{*(\varphi\varphi^\ast h)}\) and \(\Omega_{\mathrm{GW}}^{*(\psi\bar\psi h)}\), respectively. We use the same phase-transition parameters listed in Table~\ref{tab:physical_cosmological_parameters} and the same redshift prescription as in the two-body analysis. Applying Eq.~\eqref{eq:gw_redshift_to_present_epoch} then gives the corresponding present-day spectra for the scalar and fermion channels. We first examine the dependence on the collision-background parameters using the same two parameter variations as in the two-body analysis. To study the wall-boost dependence, \(v_\phi\) is fixed while \(\gamma_{\mathrm{wall}}\) is varied. The values considered and the corresponding final-state parameters are specified in the captions of Figs.~\ref{fig:scalar_three_body_gamma_wall_dependence} and \ref{fig:fermion_three_body_gamma_wall_dependence}.

\begin{figure}[t]
    \centering
    \includegraphics[width=\columnwidth]{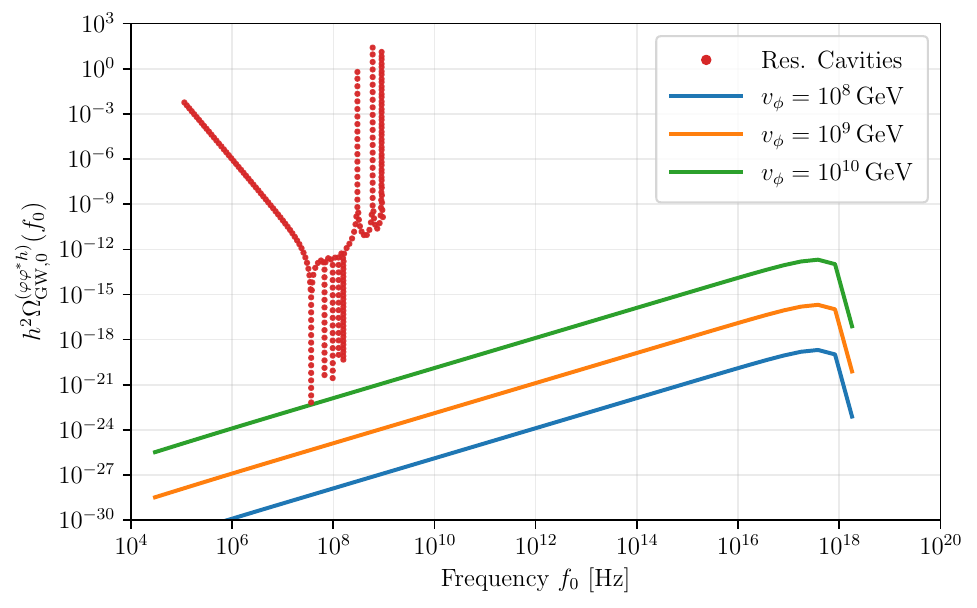}
    \caption{Present-day GW spectrum for the scalar three-body channel at fixed \(\gamma_{\mathrm{wall}}=10^7\) and \(v_\phi=10^8,10^9,10^{10}\,\mathrm{GeV}\). For each curve the final-state mass and coupling follow \(m_\varphi=0.5v_\phi\) and \(\lambda_\varphi=0.25v_\phi\).}
    \label{fig:scalar_three_body_transition_scale_dependence}
\end{figure}

\begin{figure}[t]
    \centering
    \includegraphics[width=\columnwidth]{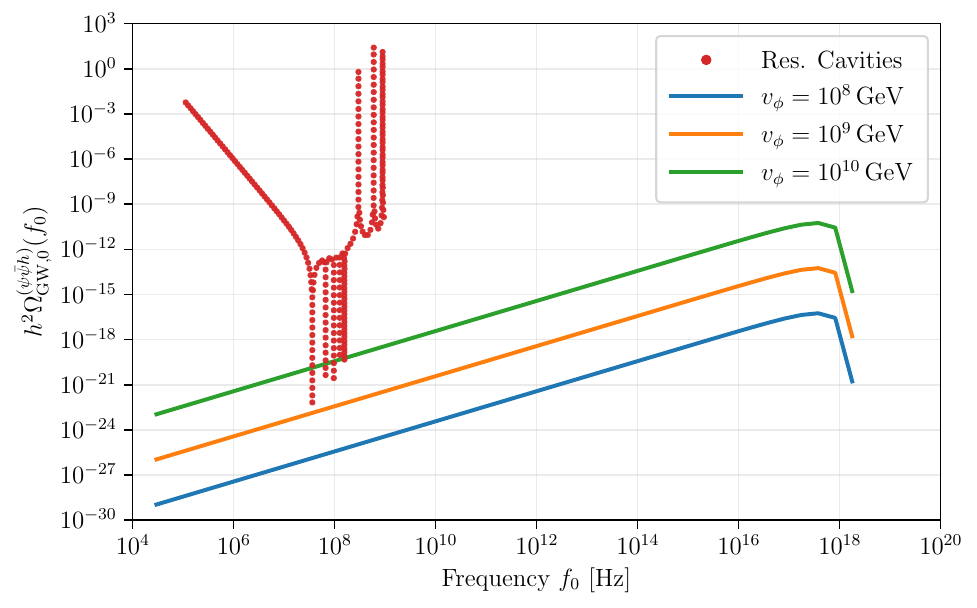}
    \caption{Present-day GW spectrum for the fermion three-body channel at fixed \(\gamma_{\mathrm{wall}}=10^7\) and \(v_\phi=10^8,10^9,10^{10}\,\mathrm{GeV}\). For each curve the final-state parameters follow \(m_\psi=0.5v_\phi\) and \(\lambda_\psi=0.5\).}
    \label{fig:fermion_three_body_transition_scale_dependence}
\end{figure}

We next fix \(\gamma_{\mathrm{wall}}\) and vary the transition scale \(v_\phi\). The values considered and the corresponding mass--coupling relations are specified in the captions of Figs.~\ref{fig:scalar_three_body_transition_scale_dependence} and \ref{fig:fermion_three_body_transition_scale_dependence}.

The background-parameter dependence of the three-body spectra follows the same qualitative behavior as in the two-body channel. At fixed \(v_\phi\), increasing \(\gamma_{\mathrm{wall}}\) raises \(\Lambda_{\max}=\gamma_{\mathrm{wall}}v_\phi\), extending the spectra to higher frequencies and increasing their amplitude. At fixed \(\gamma_{\mathrm{wall}}\), changing \(v_\phi\) mainly changes the normalization, while the characteristic present-day frequency remains nearly unchanged. This follows because the physical momentum scales with \(v_\phi\), whereas the redshift factor scales inversely with \(T_*\propto v_\phi\). These trends are the same as those found for \(\phi h\) production, showing that the dominant background-scale dependence is inherited from the collision kinematics rather than from the detailed structure of the final state.

\begin{figure}[t]
    \centering
    \includegraphics[width=\columnwidth]{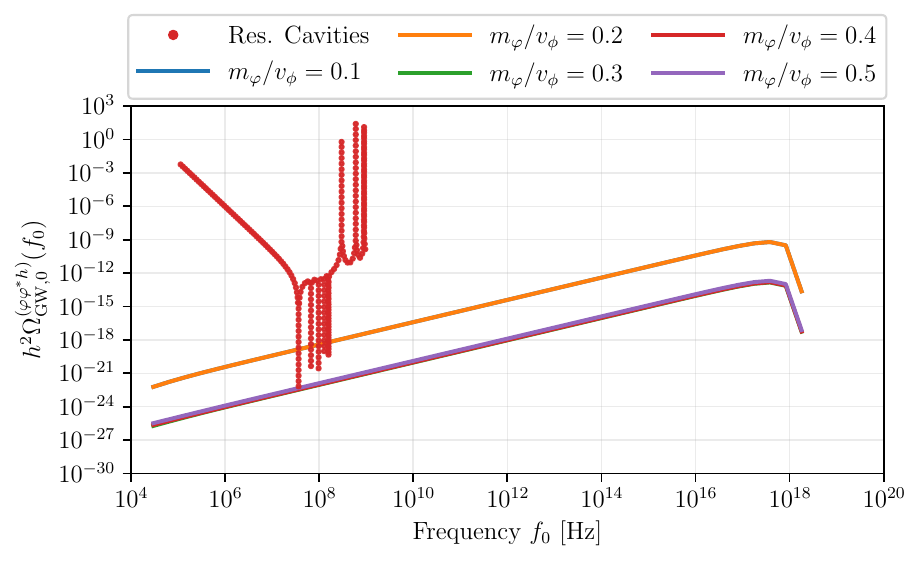}
    \caption{Mass dependence of the present-day scalar three-body GW spectrum at \(v_\phi=10^{10}\,\mathrm{GeV}\) and \(\gamma_{\mathrm{wall}}=10^7\). The five curves correspond to \(m_\varphi/v_\phi=0.1,0.2,0.3,0.4,0.5\), with \(\lambda_\varphi=m_\varphi^2/v_\phi\).}
    \label{fig:scalar_three_body_final_mass_dependence}
\end{figure}

\begin{figure}[t]
    \centering
    \includegraphics[width=\columnwidth]{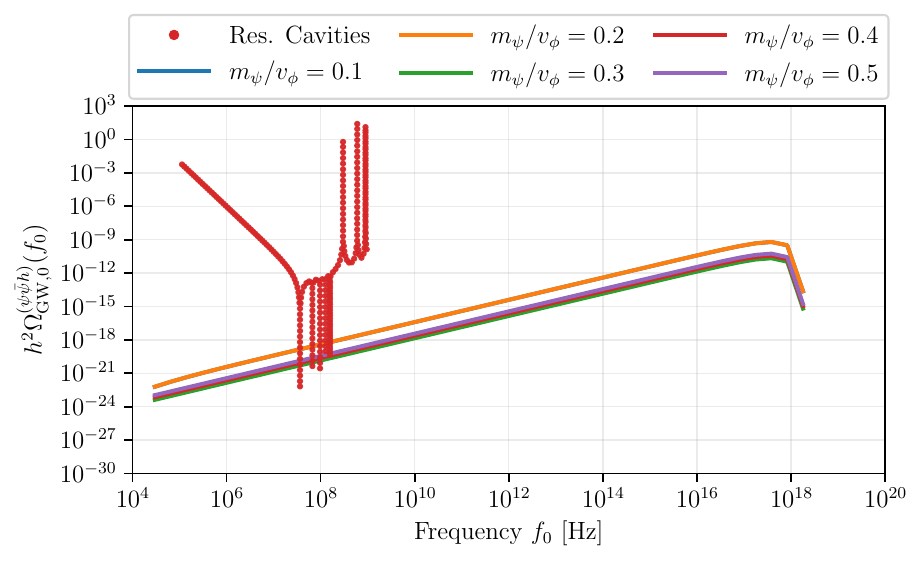}
    \caption{Mass dependence of the present-day fermion three-body GW spectrum at \(v_\phi=10^{10}\,\mathrm{GeV}\) and \(\gamma_{\mathrm{wall}}=10^7\). The five curves correspond to \(m_\psi/v_\phi=0.1,0.2,0.3,0.4,0.5\), with \(\lambda_\psi=m_\psi/v_\phi\).}
    \label{fig:fermion_three_body_final_mass_dependence}
\end{figure}

We then isolate the dependence on the final-state masses and couplings by fixing \(v_\phi=10^{10}\,\mathrm{GeV}\) and \(\gamma_{\mathrm{wall}}=10^7\), and considering
\begin{equation}
    \frac{m_\varphi}{v_\phi}=\frac{m_\psi}{v_\phi}=0.1,\,0.2,\,0.3,\,0.4,\,0.5,
    \label{eq:three_body_final_mass_scan}
\end{equation}
with \(\lambda_\varphi=m_\varphi^2/v_\phi\) and \(\lambda_\psi=m_\psi/v_\phi\). The resulting scalar and fermion mass dependences are shown in Figs.~\ref{fig:scalar_three_body_final_mass_dependence} and \ref{fig:fermion_three_body_final_mass_dependence}.

The mass parameter \(m_\phi\) enters the internal \(\phi\) propagator of both three-body channels. Since
\begin{equation}
    D_l=\frac{m_\phi^2-s_{X\bar X}}{2},
    \qquad
    s_{X\bar X}\equiv(p+q)^2,
    \label{eq:intermediate_phi_resonance_condition}
\end{equation}
the internal propagator can reach its pole inside the physical pair phase space when \(m_\phi>2m_X\). The analytic expressions above are written in the zero-width limit. For the numerical evaluation of the three-body spectra, the pole region is regulated by the replacement
\begin{equation}
    D_l\,\longrightarrow\,D_l-\frac{i}{2}m_\phi\Gamma_\phi,
    \label{eq:intermediate_phi_finite_width_regulator}
\end{equation}
with the tree-level widths
\begin{equation}
    \Gamma_\phi^{(\varphi)}=\frac{\lambda_\varphi^2}{16\pi m_\phi}\sqrt{1-\frac{4m_\varphi^2}{m_\phi^2}},\qquad \Gamma_\phi^{(\psi)}=\frac{\lambda_\psi^2m_\phi}{8\pi}\left(1-\frac{4m_\psi^2}{m_\phi^2}\right)^{3/2}.
\label{eq:intermediate_phi_partial_widths}
\end{equation}
for \(m_\phi>2m_X\), and \(\Gamma_\phi=0\) below threshold. This prescription is used only to treat the resonant propagator region in the numerical spectra. Equation~\eqref{eq:scalar_three_body_propagator_denominators} and the corresponding fermion expressions are recovered as \(\Gamma_\phi\to0\).

\begin{figure}[t]
    \centering
    \includegraphics[width=\columnwidth]{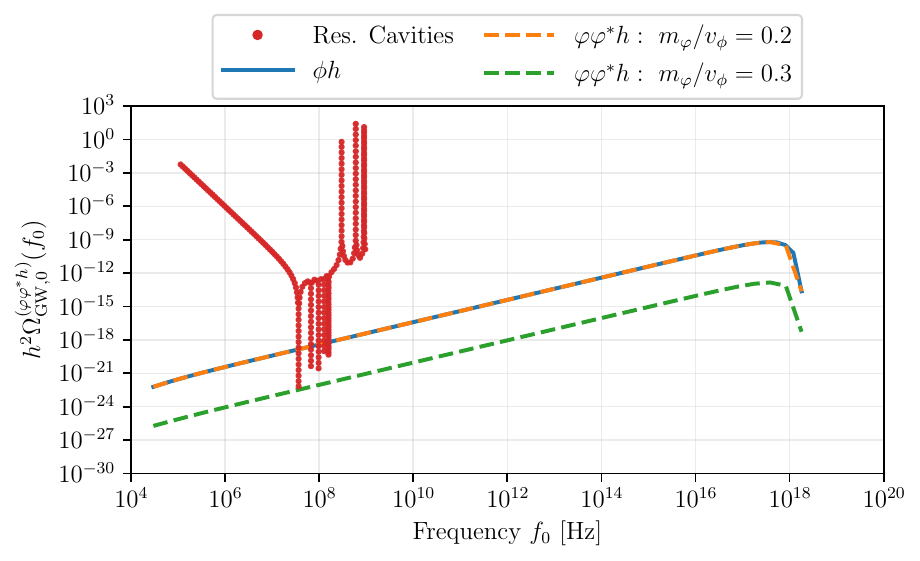}
    \caption{Comparison of the two-body \(\phi h\) spectrum with the scalar three-body \(\varphi\varphi^\ast h\) spectrum at \(v_\phi=10^{10}\,\mathrm{GeV}\) and \(\gamma_{\mathrm{wall}}=10^7\). The orange dashed curve has \(m_\varphi=2\times10^9\,\mathrm{GeV}\) and \(\lambda_\varphi=4\times10^8\,\mathrm{GeV}\), for which the on-shell internal-\(\phi\) channel is open. The green dashed curve has \(m_\varphi=3\times10^9\,\mathrm{GeV}\) and \(\lambda_\varphi=9\times10^8\,\mathrm{GeV}\), for which it is closed.}
    \label{fig:scalar_three_body_resonance_comparison}
\end{figure}

\begin{figure}[t]
    \centering
    \includegraphics[width=\columnwidth]{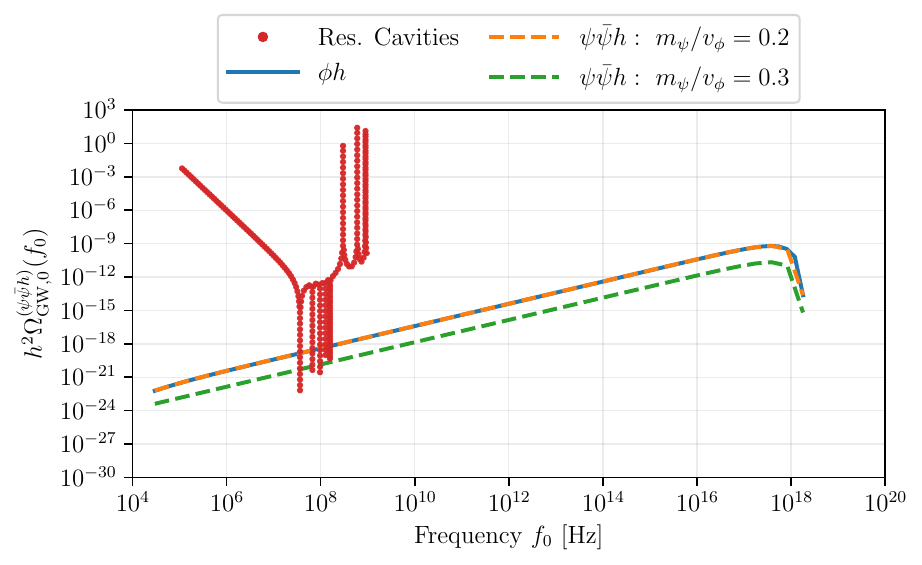}
    \caption{Comparison of the two-body \(\phi h\) spectrum with the fermion three-body \(\psi\bar\psi h\) spectrum at \(v_\phi=10^{10}\,\mathrm{GeV}\) and \(\gamma_{\mathrm{wall}}=10^7\). The orange dashed curve has \(m_\psi=2\times10^9\,\mathrm{GeV}\) and \(\lambda_\psi=0.2\), for which the on-shell internal-\(\phi\) channel is open. The green dashed curve has \(m_\psi=3\times10^9\,\mathrm{GeV}\) and \(\lambda_\psi=0.3\), for which it is closed.}
    \label{fig:fermion_three_body_resonance_comparison}
\end{figure}

The pronounced mass dependence in these results is associated with the opening or closing of the on-shell internal \(\phi\) channel. For the present benchmark, \(m_\phi=0.5v_\phi=5\times10^9\,\mathrm{GeV}\). The choices \(m_X=1\times10^9\,\mathrm{GeV}\) and \(2\times10^9\,\mathrm{GeV}\) satisfy \(m_\phi>2m_X\), so the pole at \(s_{X\bar X}=m_\phi^2\) lies inside the physical pair phase space and the corresponding three-body spectrum receives a resonant enhancement. In contrast, for \(m_X=3\times10^9\,\mathrm{GeV}\), \(4\times10^9\,\mathrm{GeV}\), and \(5\times10^9\,\mathrm{GeV}\), the condition \(m_\phi>2m_X\) is not satisfied, the pole is outside the physical region, and the spectrum is purely off shell. The change between the second and third mass points therefore directly traces the on-shell threshold rather than a smooth kinematic suppression with increasing mass. To make this interpretation explicit, Figs.~\ref{fig:scalar_three_body_resonance_comparison} and \ref{fig:fermion_three_body_resonance_comparison} compare the two-body \(\phi h\) spectrum with one three-body point on each side of the threshold. We choose \(m_X=2\times10^9\,\mathrm{GeV}\), for which the on-shell channel is open, and \(m_X=3\times10^9\,\mathrm{GeV}\), for which it is closed.

When the on-shell channel is open, the resonant part of the three-body process can be understood as production of an on-shell \(\phi\) together with the graviton, followed by \(\phi\to X\bar X\). In the independent single-channel benchmarks used here, the same partial width that appears in the regulated propagator also sets the decay of the on-shell intermediate state, so the resonant three-body spectrum follows the two-body \(\phi h\) result closely. This behavior is visible in both comparison figures: the open-channel three-body curve nearly coincides with the corresponding two-body spectrum, whereas the closed-channel curve remains suppressed. The agreement therefore provides a direct numerical check that the resonant enhancement of the three-body calculation is consistent with the two-body result when the internal \(\phi\) can become on shell.

\section{Conclusion and discussion}\label{sec:conclusion}

In this work, we have investigated a microscopic source of GWs arising from the associated production of particles and gravitons during bubble collisions in a cosmological FOPT. Unlike conventional GW production from the macroscopic dynamics of bubble walls, sound waves, and magnetohydrodynamic turbulence, this mechanism generates gravitons directly during particle production induced by the collision. Focusing on the runaway regime, we have calculated the differential graviton spectra for the two-body scalar channel and the three-body channels involving scalar or fermion particle-antiparticle pairs. We have derived analytical approximations, numerically evaluated the resulting GW spectra, and obtained their present-day counterparts by accounting for cosmological redshift.

Our results demonstrate that the microscopic GW spectra are controlled by both the kinematics of the produced particles and the characteristic scales of the bubble collisions. In particular, the bubble radius and Lorentz-contracted wall thickness determine the momentum range of the collision background that contributes to particle production. Increasing the bubble wall Lorentz factor extends the ultraviolet cutoff of the background toward higher momenta, thereby broadening the accessible phase space and extending the GW spectra to higher frequencies. The resulting high-frequency component provides a potentially distinctive signature of relativistic bubble collisions.

We have also investigated the dependence of GW spectra on the collision parameters and particle masses. For the two-body channel, increasing the wall Lorentz factor enlarges the ultraviolet scale \(\Lambda_{\max}=\gamma_{\mathrm{wall}}v_\phi\), extending the spectrum toward higher frequencies. At fixed \(\gamma_{\mathrm{wall}}=10^7\), varying \(v_\phi\) mainly changes the overall normalization, while the characteristic present-day ultraviolet frequency remains approximately unchanged under the benchmark relation \(T_*=0.1v_\phi\). The approximately linear rise of the two-body spectrum in the intermediate momentum region is also consistent with the analytical asymptotic behavior derived from the exact kinematics. In contrast, the three-body channels exhibit a pronounced dependence on the masses of the produced particle pairs. When the intermediate scalar can become on shell, \(m_\phi>2m_X\), its propagator pole enters the kinematically accessible phase space. Within the finite width prescription adopted in our numerical analysis, this leads to a resonant enhancement of the three-body spectrum, while the contribution remains suppressed when the on-shell channel is closed. These features illustrate how the microscopic GW spectra depend on both the characteristic scales of the collision background and the particle content and interactions involved in the production processes.

An important implication of our study is that microscopic associated production of particles and gravitons can provide an additional high-frequency GW component alongside the conventional signals from the same FOPT. Since the characteristic frequencies of macroscopic GW sources are primarily determined by the typical bubble separation and transition timescale, whereas the microscopic contribution can extend to frequencies associated with the Lorentz-contracted wall thickness, the two mechanisms may generate signals in widely separated frequency bands. This opens the possibility of multiband GW observations that probe both the macroscopic dynamics of the phase transition and the microscopic interactions responsible for particle production. In particular, if the collision background couples to dark matter or other hidden sector particles, their associated production with gravitons establishes a direct connection between nonthermal particle production and high-frequency GW signatures. Our results therefore identify an additional physics target for future high-frequency GW experiments.

Several aspects of the calculation merit further investigation. The collision background has been described using an approximate Fourier window, whose infrared and ultraviolet boundaries are motivated by the characteristic bubble separation and Lorentz-contracted wall thickness, respectively. A more realistic treatment of the finite collision geometry, wall dynamics, and plasma friction would improve the determination of the background spectrum and its dependence on the phase transition parameters. Furthermore, a consistent treatment of resonant intermediate states, particle interactions, and the subsequent cosmological evolution is needed for more precise predictions of the three-body contributions and their observational prospects. Assessing the detectability of the resulting high-frequency signals will also require a dedicated comparison with the sensitivities of proposed GW experiments.

Our study establishes the associated production of particles and gravitons during bubble collisions as an additional microscopic mechanism for GW generation in cosmological FOPTs. Besides the numerical GW spectra, we also provide the approximated analytic GW formula, which is convenient for future studies in concrete phase transition models. By connecting high-frequency gravitational radiation to nonthermal particle production, this mechanism provides a complementary probe of the particle content and interactions of the early Universe, with potential implications for dark matter physics and multiband GW astronomy.

\begin{acknowledgments}
We thank Siyu Jiang and Quan Chen for helpful discussions in the early stage of this work.
This work is supported by the National Natural Science Foundation of China (NNSFC) Grant No.12475111 and the Fundamental Research Funds for the Central Universities, Sun Yat-sen University.
\end{acknowledgments}

\appendix

\section{Collision background and effective action description of particle production}
\label{app:particle_production_formalism}

\subsection{Planar collision background and physical window}
\label{subsec:app_planar_momentum_window}

As the bubbles expand, their radii increase, while the relativistic motion of the walls reduces their thickness in the collision frame through Lorentz contraction. We consider an elastic collision of ultrarelativistic thin walls. The scalar potential determines the wall thickness in its rest frame. For an order-of-magnitude estimate, we take \(l_{{\mathrm{wall}},0}\sim v_\phi^{-1}\). The contracted wall thickness and the characteristic bubble radius at collision are
\begin{equation}
    l_{\mathrm{wall}}=\frac{l_{{\mathrm{wall}},0}}{\gamma_{\mathrm{wall}}}\sim\frac{1}{\gamma_{\mathrm{wall}}v_\phi},\qquad R_*\simeq(8\pi)^{1/3}v_{\mathrm{wall}}\beta^{-1}\simeq(8\pi)^{1/3}\beta^{-1}.
\label{eq:app_bubble_radius_wall_thickness_scales}
\end{equation}
where \(v_{\mathrm{wall}}\simeq1\) is used for the last expression~\cite{Caprini:2019egz,Falkowski:2012fb,Shakya:2023pp}. Thus the physical ultraviolet scale is estimated as \(l_{\mathrm{wall}}^{-1}\sim\gamma_{\mathrm{wall}}v_\phi\). For the numerical integrations, however, we adopt definite cutoff values
\begin{equation}
    \Lambda_{\min}=R_*^{-1},
    \qquad
    \Lambda_{\max}=\gamma_{\mathrm{wall}}v_\phi,
    \label{eq:app_background_ir_uv_cutoffs}
\end{equation}
so that the integration limits are fixed once the benchmark parameters are specified.

When \(R_*\) is much larger than \(l_{\mathrm{wall}}\), each collision region can be treated locally as an interaction between two planar wall segments. The corresponding hierarchy between the bubble radius and the contracted wall thickness requires parametrically
\begin{equation}
    \frac{R_*}{l_{\mathrm{wall}}}
    \sim
    (8\pi)^{1/3}\frac{\gamma_{\mathrm{wall}}v_\phi}{\beta}
    \gg1,
    \label{eq:app_planar_scale_hierarchy}
\end{equation}
which corresponds to the background-scale hierarchy \(\Lambda_{\min}\ll\Lambda_{\max}\) within the prescription above. If this hierarchy is not satisfied, curvature and the finite multi-bubble collision geometry cannot be neglected.

At length scales larger than \(R_*\), the presence and collisions of multiple bubbles modify the efficiency factor obtained from an isolated two-bubble collision~\cite{Mansour:2023fwj,Shakya:2023pp}. We therefore use \(R_*^{-1}\) as the infrared scale of the collision background and adopt \(E,\ell\geq\Lambda_{\min}\) as a modeling prescription for its Fourier modes. The finite wall thickness suppresses background components at scales above \(l_{\mathrm{wall}}^{-1}\), which we approximate by \(E,\ell\leq\Lambda_{\max}\).

Taking the collision axis along the \(z\) direction, translational invariance parallel to the walls implies that a Fourier component of the collision background carries four-momentum
\begin{equation}
    l^\mu=(E,\boldsymbol{\ell})=(E,0,0,\ell),
    \qquad
    \chi=E^2-\ell^2.
    \label{eq:app_collision_background_virtuality}
\end{equation}
The background component \(\phi(l)\) is not an incoming asymptotic particle and is not constrained by \(\chi=m_\phi^2\).

\subsection{Particle production factorization}
\label{subsec:app_effective_action}

We now derive Eq.~\eqref{eq:graviton_yield_per_collision_area} following the effective action description of particle production developed by Watkins and Widrow and the thin-wall normalization of Falkowski and No~\cite{Watkins:1991zt,Falkowski:2012fb}. For a classical collision background \(\phi(x)\), the vacuum persistence amplitude is
\begin{equation}
    \langle 0_{\mathrm{out}}|0_{\mathrm{in}}\rangle_{\phi}
    =\exp\!\left[i\Gamma[\phi]\right],
\end{equation}
and the probability of particle production is
\begin{equation}
    P_{\mathrm{prod}}
    =1-\exp\!\left[-2\,\operatorname{Im}\Gamma[\phi]\right]
    \simeq 2\,\operatorname{Im}\Gamma[\phi],
    \label{eq:app_vacuum_persistence_effective_action}
\end{equation}
where the last approximation holds for \(2\,\operatorname{Im}\Gamma[\phi]\ll1\). Keeping the quadratic term in the background field gives
\begin{equation}
    \Gamma[\phi]=\frac12\int d^4x_1\,d^4x_2\,\phi(x_1)\Gamma^{(2)}(x_1-x_2)\phi(x_2)+\mathcal O(\phi^3).
\label{eq:app_quadratic_effective_action}
\end{equation}
The Fourier conventions for the background field and the two-point function are
\begin{equation}
    \widetilde\phi(l)\equiv\int d^4x\,e^{il\cdot x}\phi(x),\qquad \widetilde\Gamma^{(2)}(E,\ell)\equiv\left.\int d^4x\,e^{il\cdot x}\Gamma^{(2)}(x)\right|_{l^\mu=(E,\boldsymbol\ell)=(E,0,0,\ell)}.
\label{eq:app_background_fourier_convention}
\end{equation}
Eqs.~\eqref{eq:app_vacuum_persistence_effective_action}--\eqref{eq:app_background_fourier_convention} express the production probability as an integral of the classical background weight \(|\widetilde\phi(E,\ell)|^2\) multiplied by \(\operatorname{Im}\widetilde\Gamma^{(2)}(E,\ell)\). This is the origin of the factorized structure in Eq.~\eqref{eq:graviton_yield_per_collision_area}. The Fourier amplitude \(\widetilde\phi(E,\ell)\) belongs to the classical source and should not be interpreted as a number density of physical \(\phi\) particles.

\begin{figure}[t]
    \centering
    \includegraphics[width=0.62\columnwidth]{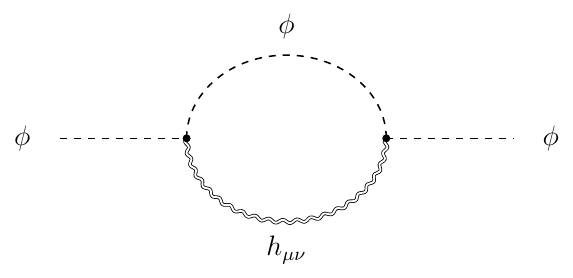}
    \caption{One-loop contribution to the background-field two-point function whose cut corresponds to the two-body final state $\phi(p)+h_{\mu\nu}(k)$. The external momentum $l$ is carried by the collision-background Fourier mode.}
    \label{fig:app_two_body_background_self_energy}
\end{figure}

\begin{figure}[t]
    \centering
    \includegraphics[width=0.96\columnwidth]{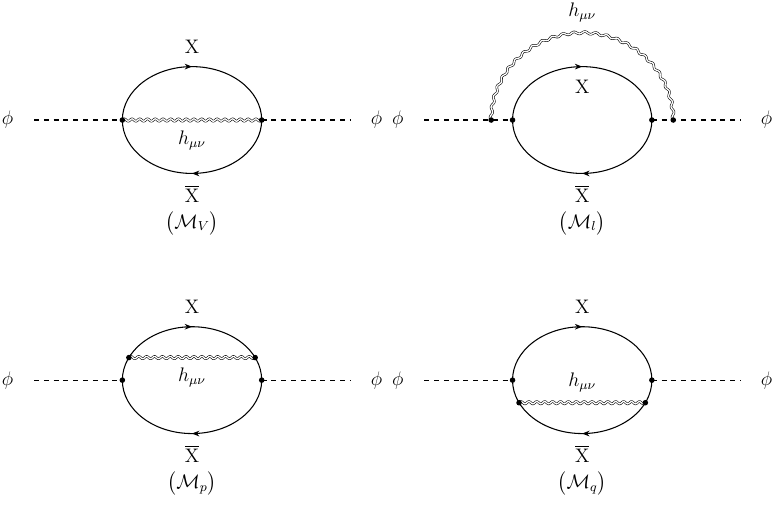}
    \caption{Two-loop contributions to the background-field two-point function whose cuts correspond to the three-body final state $X(p)+\bar X(q)+h_{\mu\nu}(k)$. The four panels correspond to $\mathcal M_{V}$, $\mathcal M_{l}$, $\mathcal M_{p}$, and $\mathcal M_{q}$, respectively, with the external momentum $l$ supplied by the collision-background Fourier mode.}
    \label{fig:app_three_body_background_self_energy}
\end{figure}

The one-loop contribution to the \(\phi\) two-point function associated with \(\phi(l)\to\phi(p)+h_{\mu\nu}(k)\) is shown in Fig.~\ref{fig:app_two_body_background_self_energy}. The corresponding two-loop topologies for \(\phi(l)\to X(p)+\bar X(q)+h_{\mu\nu}(k)\) are shown in Fig.~\ref{fig:app_three_body_background_self_energy}. The four panels match the amplitudes \(\mathcal M_{X\bar Xh,V}^{(\lambda)}\), \(\mathcal M_{X\bar Xh,l}^{(\lambda)}\), \(\mathcal M_{X\bar Xh,p}^{(\lambda)}\), and \(\mathcal M_{X\bar Xh,q}^{(\lambda)}\), respectively. In each diagram, the off-shell external line represents a Fourier component of the classical collision background, while the on-shell lines identify the physical final states.

For planar walls moving along the $z$ direction, $\phi(x)=\phi(z,t)$ and translational invariance in the transverse plane gives
\begin{equation}
    \widetilde\phi(l)
    =(2\pi)^2\delta(l_x)\delta(l_y)
    \widetilde\phi(E,\ell),
    \qquad l^\mu=(E,\boldsymbol{\ell})=(E,0,0,\ell).
    \label{eq:app_planar_fourier_delta_structure}
\end{equation}
Regulating the transverse directions by lengths $L_x$ and $L_y$ and using
\begin{equation}
    \big[(2\pi)\delta(l_i)\big]^2
    =(2\pi)L_i\delta(l_i),
    \qquad A=L_xL_y,
\end{equation}
removes the squared transverse delta functions. Using the planar particle-production normalization of Refs.~\cite{Watkins:1991zt,Falkowski:2012fb}, before imposing the background window we write the number per unit area as
\begin{equation}
    \frac{N}{A}
    =2\int_{-\infty}^{\infty}dE
    \int_{-\infty}^{\infty}\frac{d\ell}{(2\pi)^2}
    \left|\widetilde\phi(E,\ell)\right|^2
    \operatorname{Im}\!\left[
    \widetilde\Gamma^{(2)}(E,\ell)
    \right].
    \label{eq:app_particle_yield_per_area}
\end{equation}
This has the form of Eq.~(5.8) in Ref.~\cite{Watkins:1991zt} and Eq.~(2.15) in Ref.~\cite{Falkowski:2012fb}. For the channels considered here, each counted final state contains one graviton.

For a perfectly elastic collision in the ultrarelativistic, infinitely thin-wall limit, the nontrivial part of the wall profile may be written as
\begin{equation}
    \phi(z,t)
    =v_\phi\,\Theta\!\left(|z|-|t|\right).
    \label{eq:app_thin_wall_profile}
\end{equation}
Terms supported only at $E=\ell=0$ do not contribute above a physical production threshold. The remaining Fourier transform is~\cite{Falkowski:2012fb}
\begin{equation}
    \widetilde\phi(E,\ell)
    =\frac{4v_\phi}{E^2-\ell^2}
    =\frac{4v_\phi}{\chi}.
    \label{eq:app_thin_wall_fourier_transform}
\end{equation}
For the reflection-symmetric wall profile used here, the background weight is even in both $E$ and $\ell$. With the corresponding even continuation of the inclusive imaginary part in Eq.~\eqref{eq:app_particle_yield_per_area}, the integrations can be restricted to the positive quadrant with a factor of four. Imposing the physical window introduced in Sec.~\ref{sec:collision_framework} then gives
\begin{equation}
\begin{aligned}
    \frac{N}{A}
    &={}
    8\int_{\Lambda_{\min}}^{\Lambda_{\max}} dE
    \int_{\Lambda_{\min}}^{\Lambda_{\max}}\frac{d\ell}{(2\pi)^2}
    \frac{16v_\phi^2}{\chi^2}
    \operatorname{Im}\!\left[
    \widetilde\Gamma^{(2)}(E,\ell)
    \right]
    \\
    &={}
    \frac{32v_\phi^2}{\pi^2}
    \int_{\Lambda_{\min}}^{\Lambda_{\max}} dE
    \int_{\Lambda_{\min}}^{\Lambda_{\max}} d\ell\,
    \frac{\operatorname{Im}[\widetilde\Gamma^{(2)}(E,\ell)]}{(E^2-\ell^2)^2}.
    \label{eq:app_thin_wall_particle_yield}
\end{aligned}
\end{equation}
Comparison with Eq.~\eqref{eq:graviton_yield_per_collision_area} identifies
\begin{equation}
    f_{\mathrm{coll}}(E,\ell)
    =\frac{32v_\phi^2}{\pi^2\chi^2},
    \qquad
    \Lambda_{\min}\leq E,\ell\leq\Lambda_{\max},
\end{equation}
with \(f_{\mathrm{coll}}=0\) outside this interval. This is equivalent to the step-function form in Eq.~\eqref{eq:thin_wall_collision_efficiency}. The upper and lower limits are therefore properties of the collision background model.

\section{Graviton Feynman rules and external polarizations}\label{app:graviton_vertices}

This appendix collects only the graviton vertices and external-polarization relations used in the production amplitudes~\cite{Nakayama:2018pt,Barman:2023ymn,Gleisberg:2003}. All momenta in the vertex rules are taken to be incoming. An outgoing momentum in a production amplitude therefore enters the corresponding vertex with the opposite sign.

\subsection{Scalar-scalar-graviton vertex}\label{app:scalar_graviton_vertex}

For a real or complex scalar of mass $m_\varphi$, the energy-momentum tensor gives the scalar-scalar-graviton vertex
\begin{equation}
    h_{\mu\nu}(k)\,\varphi(l)\,\varphi^\ast(p):\qquad i\mathcal V^{(0)\,\mu\nu}(l,p)=-\frac{i}{M_{\mathrm{pl}}}\left[l^\mu p^\nu+l^\nu p^\mu-\eta^{\mu\nu}(l\cdot p-m_\varphi^2)\right].
\label{eq:app_scalar_scalar_graviton_vertex}
\end{equation}
\subsection{Fermion-fermion-graviton vertex}\label{app:fermion_graviton_vertex}

For a Dirac fermion of mass $m_\psi$, the fermion-fermion-graviton vertex can be written as
\begin{equation}
\begin{aligned}
    h_{\mu\nu}\,\bar\psi(p)\,\psi(q):\qquad i\mathcal V^{(1/2)\,\mu\nu}(p,q)
    &=-\frac{i}{4M_{\mathrm{pl}}}\left[\gamma^\mu(p+q)^\nu+\gamma^\nu(p+q)^\mu\right.\\
    &\left.\hspace{11em}-2\eta^{\mu\nu}(\slashed p+\slashed q-2m_\psi)\right].
\end{aligned}
\label{eq:app_fermion_fermion_graviton_vertex}
\end{equation}
\subsection{Contact vertices with one graviton}\label{app:graviton_contact_vertices}

The metric determinant in the interaction Lagrangian generates contact vertices involving the classical scalar background, the two matter fields, and a graviton. In the normalization used here, the relevant terms are
\begin{align}
    h_{\mu\nu}\phi\varphi^\ast\varphi:
    \qquad
    &-\frac{i\lambda_\varphi}{M_{\mathrm{pl}}}\eta^{\mu\nu},
    \\
    h_{\mu\nu}\phi\bar{\psi}\psi:
    \qquad
    &-\frac{i\lambda_\psi}{M_{\mathrm{pl}}}\eta^{\mu\nu}.
    \label{eq:app_contact_vertices}
\end{align}

\subsection{External graviton polarizations and polarization sum}\label{app:graviton_polarization_sum}

For a physical external graviton with momentum $k^\mu$, the polarization tensor obeys
\begin{equation}
    k_\mu k^\mu=0,\qquad k^\mu\epsilon^{(\lambda)}_{\mu\nu}(k)=0,\qquad \eta^{\mu\nu}\epsilon^{(\lambda)}_{\mu\nu}(k)=0,\qquad \epsilon^{(\lambda)}_{\mu\nu}(k)=\epsilon^{(\lambda)}_{\nu\mu}(k).
\label{eq:app_graviton_tt_conditions}
\end{equation}
For an outgoing graviton, the amplitudes are contracted with the complex-conjugate tensor $\epsilon^{(\lambda)*}_{\mu\nu}(k)$. The sum over the two physical graviton polarizations is
\begin{equation}
\begin{aligned}
\Pi_{\mu\nu,\alpha\beta}(k)
&\equiv\sum_{\lambda}
\epsilon^{(\lambda)*}_{\mu\nu}(k)
\epsilon^{(\lambda)}_{\alpha\beta}(k)\\
&=\frac12\Bigl(
\hat\eta_{\mu\alpha}\hat\eta_{\nu\beta}
+\hat\eta_{\mu\beta}\hat\eta_{\nu\alpha}
-\hat\eta_{\mu\nu}\hat\eta_{\alpha\beta}\Bigr).
\end{aligned}
\label{eq:app_graviton_polarization_sum}
\end{equation}
where
\begin{equation}
    \hat{\eta}_{\mu\nu}
    =
    \eta_{\mu\nu}
    -
    \frac{k_\mu\bar{k}_\nu+\bar{k}_\mu k_\nu}{k\cdot\bar{k}},
    \qquad
    \bar{k}^\mu=(k,-\boldsymbol{k}).
    \label{eq:app_hat_eta_definition}
\end{equation}

\section{Linear asymptotic behavior of the two-body spectrum}
\label{app:two_body_asymptotic}

Here we derive the approximately linear behavior in Eq.~\eqref{eq:two_body_linear_momentum_scaling} from the exact two-body kinematics. Starting from Eq.~\eqref{eq:two_body_gw_spectrum_kinematic_limits}, we define
\begin{equation}
    \mathcal I_m(E,k)
    \equiv
    \int_{\ell_-}^{\ell_+}d\ell\,\ell^3
    \frac{(1-\mu_0^2)^2}{(E^2-\ell^2)^2},
    \label{eq:app_two_body_longitudinal_inner_integral}
\end{equation}
where
\begin{equation}
    \Delta(E,k)=\sqrt{(E-k)^2-m_\phi^2},
    \qquad
    \ell_+=k+\Delta(E,k),
    \qquad
    \ell_-=\max\!\left\{\Lambda_{\min},\,|k-\Delta(E,k)|\right\},
    \label{eq:app_two_body_exact_longitudinal_limits}
\end{equation}
and the outer integration begins at \(E_{\min}(k)\) in Eq.~\eqref{eq:two_body_minimum_background_energy}.
The inner integral can be evaluated analytically using the exact kinematic boundaries. For convenience, we define
\begin{equation}
\begin{aligned}
    \mathcal F(E,\ell,k,m_\phi)={}&
    \frac{\ell^4}{64k^4}
    +\frac{\ell^2\left[-(E-2k)^2+2m_\phi^2\right]}{16k^4}
    +\frac{m_\phi^4(4Ek+m_\phi^2)^2}
    {32E^2k^4(E^2-\ell^2)}
    \\
    &+\frac{m_\phi^2(4Ek+m_\phi^2)
    \left[4E(E-k)(m_\phi^2+2Ek)-m_\phi^4\right]}
    {32E^4k^4}
    \ln\!\left(1-\frac{\ell^2}{E^2}\right)
    \\
    &+\frac{(-E^2+2Ek+m_\phi^2)^4}{16E^4k^4}
    \ln\!\left(\frac{\ell}{E}\right),
\end{aligned}
\label{eq:app_two_body_longitudinal_antiderivative}
\end{equation}
so that
\begin{equation}
    \mathcal I_m(E,k)
    =\mathcal F(E,\ell_+,k,m_\phi)
    -\mathcal F(E,\ell_-,k,m_\phi).
    \label{eq:app_two_body_exact_inner_integral}
\end{equation}
The logarithms have been written with dimensionless arguments, additive constants in the antiderivative cancel between the two limits. Eq.~\eqref{eq:app_two_body_exact_inner_integral} retains the exact kinematic boundaries and remains regular before the asymptotic expansion.

To isolate the approximately linear regime after the exact \(\ell\) integration, we introduce the two small expansion parameters
\begin{equation}
    \epsilon\equiv\frac{k}{\Lambda_{\max}},
    \qquad
    \delta_m\equiv\frac{m_\phi^2}{k\Lambda_{\max}}.
    \label{eq:app_two_body_asymptotic_expansion_parameters}
\end{equation}
Here \(\epsilon\) measures the emitted graviton momentum relative to the ultraviolet cutoff, while \(\delta_m\) measures the leading mass-dependent kinematic suppression.

In the ultraviolet-dominated part of the \(E\) integral, \(E\gg k\), expansion of Eq.~\eqref{eq:app_two_body_exact_inner_integral} gives
\begin{equation}
    \mathcal I_m(E,k)
    =\mathcal I_0(E,k)
    +\frac{m_\phi^2}{k^2}
    \left[
        \ln\!\left(\frac{m_\phi^2}{4kE}\right)+2
    \right]
    +\cdots,
    \label{eq:app_two_body_high_energy_inner_expansion}
\end{equation}
where \(\mathcal I_0\) denotes the leading high-energy contribution to the inner integral. The omitted terms in Eq.~\eqref{eq:app_two_body_high_energy_inner_expansion} carry an additional suppression by \(k/E\) or by \(m_\phi^2/(kE)\). Integrating over \(E\) up to \(\Lambda_{\max}\), while retaining the contribution from the lower background boundary exactly, gives
\begin{equation}
\begin{aligned}
    \int_{E_{\min}(k)}^{\Lambda_{\max}}dE\,\mathcal I_m(E,k)
    =\frac{\Lambda_{\max}^2}{3k}
    \Bigg[
        1-4\epsilon
        +\mathcal C_{\mathrm{quad}}(k)\epsilon^2
        -3\delta_m\mathcal C_{\mathrm{kin}}(k)
        +9\delta_m
        \\
        \hspace{2.7cm}
        +\mathcal O\!\left(
            \epsilon^3,
            \delta_m\epsilon\mathcal C_{\mathrm{kin}}^2(k),
            \delta_m^2\mathcal C_{\mathrm{kin}}(k)
        \right)
    \Bigg].
\end{aligned}
\label{eq:app_two_body_background_energy_expansion}
\end{equation}
The dimensionless coefficient of the quadratic momentum correction is
\begin{equation}
\mathcal C_{\mathrm{quad}}(k)=
\begin{cases}
\displaystyle \frac65\ln\!\left(\frac{\Lambda_{\max}}{k}\right)+\frac{91}{25}-\frac1{25}\left(\frac{\Lambda_{\min}}{k}\right)^5, & k\geq\Lambda_{\min},\\[6pt]
\displaystyle \frac65\ln\!\left(\frac{\Lambda_{\max}}{\Lambda_{\min}}\right)+\frac85+3\frac{\Lambda_{\min}}{k}-\left(\frac{\Lambda_{\min}}{k}\right)^2, & k\leq\Lambda_{\min},
\end{cases}
\label{eq:app_two_body_quadratic_momentum_coefficient}
\end{equation}
its dependence on \(\Lambda_{\min}\) retains the effect of the lower background scale. The kinematic logarithmic coefficient is defined by
\begin{equation}
    \mathcal C_{\mathrm{kin}}(k)
    \equiv
    \ln\!\left(\frac{4k\Lambda_{\max}}{m_\phi^2}\right)
    =\ln\!\left(\frac{4}{\delta_m}\right).
    \label{eq:app_two_body_mass_log_coefficient}
\end{equation}
The logarithmic part originates from the exact upper kinematic boundary in Eq.~\eqref{eq:app_two_body_longitudinal_antiderivative}.

Substituting Eq.~\eqref{eq:app_two_body_background_energy_expansion} into Eq.~\eqref{eq:two_body_gw_spectrum_kinematic_limits} and retaining only the leading term gives Eq.~\eqref{eq:two_body_linear_momentum_scaling}. The remaining terms quantify departures from the linear behavior. The expansion is controlled by
\begin{equation}
    \delta_m\ll1,
    \qquad
    \epsilon\ll1,
    \qquad\Longleftrightarrow\qquad
    \frac{m_\phi^2}{\Lambda_{\max}}\ll k\ll\Lambda_{\max}.
    \label{eq:app_two_body_linear_regime_validity}
\end{equation}
Within the window in Eq.~\eqref{eq:app_two_body_linear_regime_validity}, the first term in Eq.~\eqref{eq:app_two_body_background_energy_expansion} produces the linear scaling \(\Omega_{\mathrm{GW}}^{*(\phi h)}\propto k\), while the remaining terms describe the first departures from this behavior. Outside this window, the full integral in Eq.~\eqref{eq:two_body_gw_spectrum_kinematic_limits} is used.

\section{Three-body amplitudes, phase space, and spectra}\label{app:three_body_calculation}

We consider the three-body process
\begin{equation}
    \phi(l) \rightarrow X(p)+\bar X(q)+h_{\mu\nu}(k),
\end{equation}
where \(X\bar X\) denotes either the scalar pair \(\varphi\varphi^\ast\) or the fermion pair \(\psi\bar\psi\), with \(m_X\) denoting the corresponding particle mass. The graviton is denoted by \(h_{\mu\nu}(k)\), and \(\phi(l)\) is a Fourier component of the classical background. Using the same collision-frame parametrization as in Sec.~\ref{sec:three_body_emission}, we write
\begin{equation}
\begin{aligned}
    l^\mu&=(E,\boldsymbol{\ell})=(E,0,0,\ell),\\
    k^\mu&=(k,\boldsymbol{k})
    =k\bigl(1,\sigma_k\cos\theta_k,\sigma_k\sin\theta_k,\mu_k\bigr),
    \qquad
    \mu_k\equiv\widehat{\boldsymbol\ell}\cdot\widehat{\boldsymbol k},
    \qquad
    \sigma_k\equiv\sqrt{1-\mu_k^2},
    \\
    p^\mu&=(E_p,\boldsymbol{p})
    =\bigl(E_p,p\sigma_p\cos\theta_p,
    p\sigma_p\sin\theta_p,p\mu_p\bigr),
    \qquad
    \mu_p\equiv\widehat{\boldsymbol\ell}\cdot\widehat{\boldsymbol p},
    \qquad
    \sigma_p\equiv\sqrt{1-\mu_p^2},
    \\
    q^\mu&=(E_q,\boldsymbol{q})=l^\mu-p^\mu-k^\mu\\
    &=\bigl(E-k-E_p,
    -k\sigma_k\cos\theta_k-p\sigma_p\cos\theta_p,-k\sigma_k\sin\theta_k-p\sigma_p\sin\theta_p,
    \ell-k\mu_k-p\mu_p\bigr).
\end{aligned}
\label{eq:app_three_body_four_momenta}
\end{equation}
here
\begin{equation}
    l^\mu=p^\mu+q^\mu+k^\mu,\qquad \chi\equiv l_\mu l^\mu=E^2-\ell^2\neq m_\phi^2,\qquad p_\mu p^\mu=q_\mu q^\mu=m_X^2,\qquad k_\mu k^\mu=0.
\end{equation}

\subsection{Scalar-pair amplitudes}\label{app:scalar_three_body_calculation}

For the scalar pair, the matter vertex following from $-\lambda_\varphi\phi|\varphi|^2$ is $-i\lambda_\varphi$. Before any kinematic simplification, the four amplitudes are
\begin{equation}
\label{eq:scalar_four_raw_amplitudes}
\begin{aligned}
    i\mathcal M_{\varphi\varphi^\ast h,V}^{(\lambda)}
    & =\left(-\frac{i\lambda_\varphi}{M_{\mathrm{pl}}}\eta^{\mu\nu}\right)
    \epsilon^{(\lambda)*}_{\mu\nu}(k),\\[2pt]
    i\mathcal M_{\varphi\varphi^\ast h,l}^{(\lambda)}
    & =(-i\lambda_\varphi)
    \frac{i}{(l-k)^2-m_\phi^2}
    \left[i\mathcal V^{(0)\,\mu\nu}(l,k-l)\right]
    \epsilon^{(\lambda)*}_{\mu\nu}(k),\\[2pt]
    i\mathcal M_{\varphi\varphi^\ast h,p}^{(\lambda)}
    & =(-i\lambda_\varphi)
    \frac{i}{(p+k)^2-m_\varphi^2}
    \left[i\mathcal V^{(0)\,\mu\nu}(p+k,-p)\right]
    \epsilon^{(\lambda)*}_{\mu\nu}(k),\\[2pt]
    i\mathcal M_{\varphi\varphi^\ast h,q}^{(\lambda)}
    & =(-i\lambda_\varphi)
    \frac{i}{(q+k)^2-m_\varphi^2}
    \left[i\mathcal V^{(0)\,\mu\nu}(q+k,-q)\right]
    \epsilon^{(\lambda)*}_{\mu\nu}(k).
\end{aligned}
\end{equation}

We now reduce these expressions in the same order. The contact contribution vanishes immediately because the physical graviton is traceless,
\begin{equation}
    \eta^{\mu\nu}\epsilon^{(\lambda)*}_{\mu\nu}(k)=0,
    \qquad
    \mathcal M_{\varphi\varphi^\ast h,V}^{(\lambda)}=0.
\end{equation}
For the three propagators, the on-shell conditions for the final-state particles and $k^2=0$ give
\begin{equation}
\begin{aligned}
(l-k)^2-m_\phi^2
&=\chi-2l\cdot k-m_\phi^2=-2D_l,\\
D_l&\equiv l\cdot k-\frac{\chi-m_\phi^2}{2}
=k(E-\ell\mu_k)-\frac{\chi-m_\phi^2}{2},\\[2pt]
(p+k)^2-m_\varphi^2
&=2p\cdot k=2D_p,\\
D_p&\equiv p\cdot k
=k\bigl(E_p-p\,\widehat{\boldsymbol p}\!\cdot\!\widehat{\boldsymbol k}\bigr),\\[2pt]
(q+k)^2-m_\varphi^2
&=2q\cdot k=2D_q,\\
D_q&\equiv q\cdot k=l\cdot k-p\cdot k\\
&=k\bigl(E-E_p-\ell\mu_k
+p\,\widehat{\boldsymbol p}\!\cdot\!\widehat{\boldsymbol k}\bigr).
\end{aligned}
\label{eq:app_three_body_denominators}
\end{equation}
The scalar-graviton vertex in Eq.~\eqref{eq:app_scalar_scalar_graviton_vertex}, together with the tracelessness and transversality conditions in Eq.~\eqref{eq:app_graviton_tt_conditions}, yields
\begin{equation}
\begin{aligned}
\epsilon^{(\lambda)*}_{\mu\nu}
\left[i\mathcal V^{(0)\,\mu\nu}(l,k-l)\right]
&=\frac{2i}{M_{\mathrm{pl}}}
\epsilon^{(\lambda)*}_{\mu\nu}l^\mu l^\nu,\\
\epsilon^{(\lambda)*}_{\mu\nu}
\left[i\mathcal V^{(0)\,\mu\nu}(p+k,-p)\right]
&=\frac{2i}{M_{\mathrm{pl}}}
\epsilon^{(\lambda)*}_{\mu\nu}p^\mu p^\nu,\\
\epsilon^{(\lambda)*}_{\mu\nu}
\left[i\mathcal V^{(0)\,\mu\nu}(q+k,-q)\right]
&=\frac{2i}{M_{\mathrm{pl}}}
\epsilon^{(\lambda)*}_{\mu\nu}q^\mu q^\nu.
\end{aligned}
\end{equation}
Substituting these reductions into Eq.~\eqref{eq:scalar_four_raw_amplitudes} gives
\begin{equation}
\label{eq:scalar_four_reduced_amplitudes}
\begin{aligned}
    i\mathcal M_{\varphi\varphi^\ast h,V}^{(\lambda)}
    &=0,\\[2pt]
    i\mathcal M_{\varphi\varphi^\ast h,l}^{(\lambda)}
    &=-\frac{i\lambda_\varphi}{M_{\mathrm{pl}}}
    \frac{l^\mu l^\nu\epsilon^{(\lambda)*}_{\mu\nu}(k)}{D_l},\\[2pt]
    i\mathcal M_{\varphi\varphi^\ast h,p}^{(\lambda)}
    &=\frac{i\lambda_\varphi}{M_{\mathrm{pl}}}
    \frac{p^\mu p^\nu\epsilon^{(\lambda)*}_{\mu\nu}(k)}{D_p},\\[2pt]
    i\mathcal M_{\varphi\varphi^\ast h,q}^{(\lambda)}
    &=\frac{i\lambda_\varphi}{M_{\mathrm{pl}}}
    \frac{q^\mu q^\nu\epsilon^{(\lambda)*}_{\mu\nu}(k)}{D_q}.
\end{aligned}
\end{equation}
Therefore the total scalar amplitude is the sum of the three nonvanishing contributions,
\begin{equation}
\label{eq:scalar_amplitude_total_background_mode}
\begin{aligned}
    i\mathcal M_{\varphi\varphi^\ast h}^{(\lambda)}
    &\equiv i\mathcal M_{\varphi\varphi^\ast h,l}^{(\lambda)}
    +i\mathcal M_{\varphi\varphi^\ast h,p}^{(\lambda)}
    +i\mathcal M_{\varphi\varphi^\ast h,q}^{(\lambda)}\\
    &=\frac{i\lambda_\varphi}{M_{\mathrm{pl}}}\epsilon^{(\lambda)*}_{\mu\nu}(k)
    \left[
    -\frac{l^\mu l^\nu}{D_l}
    +\frac{p^\mu p^\nu}{D_p}
    +\frac{q^\mu q^\nu}{D_q}
    \right].
\end{aligned}
\end{equation}

To obtain the squared amplitude, we next carry out the polarization sum. For two generic four-vectors $a^\mu$ and $b^\mu$, Eq.~\eqref{eq:app_graviton_polarization_sum} gives
\begin{equation}
\begin{aligned}
    &\sum_\lambda
    \left(\epsilon^{(\lambda)*}_{\mu\nu}a^\mu a^\nu\right)
    \left(\epsilon^{(\lambda)}_{\alpha\beta}b^\alpha b^\beta\right)\\
    &\qquad=
    \left(\hat\eta_{\mu\alpha}a^\mu b^\alpha\right)
    \left(\hat\eta_{\nu\beta}a^\nu b^\beta\right)
    -\frac12
    \left(\hat\eta_{\mu\nu}a^\mu a^\nu\right)
    \left(\hat\eta_{\alpha\beta}b^\alpha b^\beta\right).
\end{aligned}
\label{eq:app_generic_scalar_polarization_contraction}
\end{equation}
Define
\begin{equation}
    \mathcal T(a,b)\equiv \hat\eta_{\mu\nu}a^\mu b^\nu
    =a\cdot b
    -\frac{(a\cdot k)(b\cdot\bar k)+(a\cdot\bar k)(b\cdot k)}{k\cdot\bar k}
    =-\boldsymbol a\cdot\boldsymbol b
    +(\widehat{\boldsymbol k}\cdot\boldsymbol a)
     (\widehat{\boldsymbol k}\cdot\boldsymbol b),
\label{eq:app_transverse_product}
\end{equation}
so that $\mathcal T(k,a)=0$. The three independent transverse structures are
\begin{equation}
\begin{aligned}
    \mathcal T_1
    &\equiv \mathcal T(l,l)
    =-\ell^2(1-\mu_k^2)
    =-\ell^2\sigma_k^2,\\
    \mathcal T_2
    &\equiv \mathcal T(p,p)
    =-p^2\!\left[1-
    \left(\widehat{\boldsymbol p}\!\cdot\!\widehat{\boldsymbol k}\right)^2\right],\\
    \mathcal T_3
    &\equiv \mathcal T(l,p)
    =\ell p\!\left[
    \mu_k\left(\widehat{\boldsymbol p}\!\cdot\!\widehat{\boldsymbol k}\right)-\mu_p
    \right].
\end{aligned}
\label{eq:app_t123_definition}
\end{equation}
Because $q=l-p-k$,
\begin{equation}
    \mathcal T(q,q)=\mathcal T_1+\mathcal T_2-2\mathcal T_3,
    \qquad
    \mathcal T(l,q)=\mathcal T_1-\mathcal T_3,
    \qquad
    \mathcal T(p,q)=\mathcal T_3-\mathcal T_2.
\label{eq:app_q_transverse_relations}
\end{equation}
Applying these relations to Eq.~\eqref{eq:scalar_amplitude_total_background_mode}, the squared scalar amplitude summed over the two physical graviton polarizations reduces to
\begin{equation}
\begin{aligned}
\overline{|\mathcal M_{\varphi\varphi^\ast h}|^2}
=\left(\frac{\lambda_\varphi}{M_{\mathrm{pl}}}\right)^2
\Bigg[&\frac{\mathcal T_1^2}{2D_l^2}
+\frac{\mathcal T_2^2}{2D_p^2}
+\frac{(\mathcal T_1+\mathcal T_2-2\mathcal T_3)^2}{2D_q^2}\\
&-\frac{2\mathcal T_3^2-\mathcal T_1\mathcal T_2}{D_lD_p}
-\frac{\mathcal T_1^2-2\mathcal T_1\mathcal T_3+2\mathcal T_3^2-\mathcal T_1\mathcal T_2}{D_lD_q}\\
&+\frac{2\mathcal T_3^2-2\mathcal T_2\mathcal T_3+\mathcal T_2^2-\mathcal T_1\mathcal T_2}{D_pD_q}
\Bigg].
\end{aligned}
\label{eq:scalar_amplitude_squared_background_mode}
\end{equation}

\subsection{Fermion-pair amplitudes}\label{app:fermion_three_body_calculation}

For the fermion pair, the corresponding Feynman-rule expressions are
\begin{equation}
\label{eq:fermion_four_raw_amplitudes}
\begin{aligned}
    i\mathcal M_{\psi\bar\psi h,V}^{(\lambda)}
    & =\bar u(p)\left(-\frac{i\lambda_\psi}{M_{\mathrm{pl}}}\eta^{\mu\nu}\right)v(q)\,
    \epsilon^{(\lambda)*}_{\mu\nu}(k),\\[2pt]
    i\mathcal M_{\psi\bar\psi h,l}^{(\lambda)}
    & =\bar u(p)(-i\lambda_\psi)v(q)\,
    \frac{i}{(l-k)^2-m_\phi^2}
    \left[i\mathcal V^{(0)\,\mu\nu}(l,k-l)\right]
    \epsilon^{(\lambda)*}_{\mu\nu}(k),\\[2pt]
    i\mathcal M_{\psi\bar\psi h,p}^{(\lambda)}
    & =\bar u(p)
    \left.i\mathcal V^{(1/2)\,\mu\nu}\right|_{p\,\mathrm{leg}}
    \frac{i(\slashed p+\slashed k+m_\psi)}{(p+k)^2-m_\psi^2}
    (-i\lambda_\psi)v(q)\,
    \epsilon^{(\lambda)*}_{\mu\nu}(k),\\[2pt]
    i\mathcal M_{\psi\bar\psi h,q}^{(\lambda)}
    & =\bar u(p)(-i\lambda_\psi)
    \frac{i(-\slashed q-\slashed k+m_\psi)}{(q+k)^2-m_\psi^2}
    \left.i\mathcal V^{(1/2)\,\mu\nu}\right|_{q\,\mathrm{leg}}
    v(q)\,\epsilon^{(\lambda)*}_{\mu\nu}(k).
\end{aligned}
\end{equation}

We now simplify Eq.~\eqref{eq:fermion_four_raw_amplitudes}. Tracelessness sets the contact term to zero, while the same scalar-graviton contraction and denominator reduction used above applies to the $l$ contribution. For the fermionic terms we use the symmetry, transversality, and tracelessness of the external graviton, together with
\begin{equation}
    \bar u(p)(\slashed p-m_\psi)=0,
    \qquad
    (\slashed q+m_\psi)v(q)=0,
\end{equation}
so that
\begin{equation}
    \bar u(p)\slashed p=m_\psi\bar u(p),
    \qquad
    \slashed q\,v(q)=-m_\psi v(q),
\end{equation}
and with momentum conservation
\begin{equation}
    \slashed l=\slashed p+\slashed q+\slashed k.
\end{equation}
The contraction of the symmetrized fermion-graviton vertex may be written as
\begin{equation}
    \epsilon^{(\lambda)*}_{\mu\nu}\,
    \frac12\left(a^\mu\gamma^\nu+a^\nu\gamma^\mu\right)
    =\epsilon^{(\lambda)*}_{\mu\nu}a^\mu\gamma^\nu.
\end{equation}
Using these relations in the raw $p$-leg expression first gives
\begin{equation}
\begin{aligned}
    i\mathcal M_{\psi\bar\psi h,p}^{(\lambda)}
    =\frac{i\lambda_\psi}{2M_{\mathrm{pl}}D_p}
    \epsilon^{(\lambda)*}_{\mu\nu}p^\mu
    \bar u(p)\gamma^\nu(\slashed l+2m_\psi)v(q).
\end{aligned}
\end{equation}
The spinor chain is reduced step by step as
\begin{equation}
\begin{aligned}
\bar u(p)\gamma^\nu(\slashed l+2m_\psi)v(q)
={}&\bar u(p)\gamma^\nu
(\slashed p+\slashed q+\slashed k+2m_\psi)v(q)\\
={}&\bar u(p)\gamma^\nu(\slashed p+\slashed k+m_\psi)v(q)\\
={}&\bar u(p)\gamma^\nu\slashed p\,v(q)
+\bar u(p)\gamma^\nu\slashed k\,v(q)
+m_\psi\bar u(p)\gamma^\nu v(q).
\end{aligned}
\end{equation}
Using
\begin{equation}
    \gamma^\nu\slashed p=2p^\nu-\slashed p\gamma^\nu,
\end{equation}
we obtain
\begin{equation}
    \bar u(p)\gamma^\nu\slashed p\,v(q)
    =2p^\nu\bar u(p)v(q)-m_\psi\bar u(p)\gamma^\nu v(q),
\end{equation}
so the mass terms cancel and
\begin{equation}
    \bar u(p)\gamma^\nu(\slashed l+2m_\psi)v(q)
    =2p^\nu\bar u(p)v(q)
    +\bar u(p)\gamma^\nu\slashed k\,v(q).
\end{equation}
For the $q$ term, the same steps give
\begin{equation}
\begin{aligned}
    i\mathcal M_{\psi\bar\psi h,q}^{(\lambda)}
    =\frac{i\lambda_\psi}{2M_{\mathrm{pl}}D_q}
    \epsilon^{(\lambda)*}_{\mu\nu}q^\mu
    \bar u(p)(\slashed l-2m_\psi)\gamma^\nu v(q).
\end{aligned}
\end{equation}
Using $\bar u(p)\slashed p=m_\psi\bar u(p)$ gives
\begin{equation}
\begin{aligned}
\bar u(p)(\slashed l-2m_\psi)\gamma^\nu v(q)
={}&\bar u(p)(\slashed p+\slashed q+\slashed k-2m_\psi)\gamma^\nu v(q)\\
={}&\bar u(p)(\slashed q+\slashed k-m_\psi)\gamma^\nu v(q)\\
={}&\bar u(p)\slashed q\gamma^\nu v(q)
+\bar u(p)\slashed k\gamma^\nu v(q)
-m_\psi\bar u(p)\gamma^\nu v(q).
\end{aligned}
\end{equation}
The identity
\begin{equation}
    \slashed q\gamma^\nu=2q^\nu-\gamma^\nu\slashed q
\end{equation}
gives
\begin{equation}
    \bar u(p)\slashed q\gamma^\nu v(q)
    =2q^\nu\bar u(p)v(q)+m_\psi\bar u(p)\gamma^\nu v(q),
\end{equation}
so the mass terms again cancel,
\begin{equation}
    \bar u(p)(\slashed l-2m_\psi)\gamma^\nu v(q)
    =2q^\nu\bar u(p)v(q)+\bar u(p)\slashed k\gamma^\nu v(q).
\end{equation}
Finally, using
\begin{equation}
    \slashed k\gamma^\nu=2k^\nu-\gamma^\nu\slashed k
\end{equation}
and $k^\nu\epsilon^{(\lambda)*}_{\mu\nu}=0$, one finds
\begin{equation}
    \epsilon^{(\lambda)*}_{\mu\nu}q^\mu
    \bar u(p)\slashed k\gamma^\nu v(q)
    =-\epsilon^{(\lambda)*}_{\mu\nu}q^\mu
    \bar u(p)\gamma^\nu\slashed k\,v(q).
\end{equation}
Collecting the reduced expressions,
\begin{equation}
\label{eq:fermion_four_reduced_amplitudes}
\begin{aligned}
    i\mathcal M_{\psi\bar\psi h,V}^{(\lambda)}
    &=0,\\[2pt]
    i\mathcal M_{\psi\bar\psi h,l}^{(\lambda)}
    &=-\frac{i\lambda_\psi}{M_{\mathrm{pl}}}
    \frac{l^\mu l^\nu\epsilon^{(\lambda)*}_{\mu\nu}}{D_l}\,
    \bar u(p)v(q),\\[2pt]
    i\mathcal M_{\psi\bar\psi h,p}^{(\lambda)}
    &=\frac{i\lambda_\psi}{M_{\mathrm{pl}}}
    \epsilon^{(\lambda)*}_{\mu\nu}
    \left[
        \frac{p^\mu p^\nu}{D_p}\,\bar u(p)v(q)
        +\frac{p^\mu}{2D_p}\,
        \bar u(p)\gamma^\nu\slashed k\,v(q)
    \right],\\[2pt]
    i\mathcal M_{\psi\bar\psi h,q}^{(\lambda)}
    &=\frac{i\lambda_\psi}{M_{\mathrm{pl}}}
    \epsilon^{(\lambda)*}_{\mu\nu}
    \left[
        \frac{q^\mu q^\nu}{D_q}\,\bar u(p)v(q)
        -\frac{q^\mu}{2D_q}\,
        \bar u(p)\gamma^\nu\slashed k\,v(q)
    \right].
\end{aligned}
\end{equation}

Their sum is
\begin{equation}
\begin{aligned}
i\mathcal M_{\psi\bar\psi h}^{(\lambda)}
={}&i\mathcal M_{\psi\bar\psi h,V}^{(\lambda)}
+i\mathcal M_{\psi\bar\psi h,l}^{(\lambda)}
+i\mathcal M_{\psi\bar\psi h,p}^{(\lambda)}
+i\mathcal M_{\psi\bar\psi h,q}^{(\lambda)}\\
={}&\frac{i\lambda_\psi}{M_{\mathrm{pl}}}\,
\epsilon^{(\lambda)*}_{\mu\nu}
\Bigg\{
\left(-\frac{l^\mu l^\nu}{D_l}
+\frac{p^\mu p^\nu}{D_p}
+\frac{q^\mu q^\nu}{D_q}\right)\bar u(p)v(q)\\
&\hspace{4.5em}
+\left(\frac{p^\mu}{2D_p}-\frac{q^\mu}{2D_q}\right)
\bar u(p)\gamma^\nu\slashed k\,v(q)
\Bigg\}.
\end{aligned}
\label{eq:fermion_amplitude_total_background_mode}
\end{equation}
To square the amplitude, the required spin sums are
\begin{equation}
    \sum_{\mathrm{spin}}\left|\bar u(p)v(q)\right|^2
    =\Tr\!\left[(\slashed p+m_\psi)(\slashed q-m_\psi)\right]
    =4(p\cdot q-m_\psi^2),
\label{eq:fermion_scalar_bilinear_spin_sum}
\end{equation}
\begin{equation}
    \sum_{\mathrm{spin}}[\bar u(p)v(q)]
    [\bar u(p)\gamma^\beta\slashed k\,v(q)]^*
    =4\left[(p\cdot q-m_\psi^2)k^\beta-D_pq^\beta+D_qp^\beta\right],
\label{eq:fermion_mixed_bilinear_spin_sum}
\end{equation}
and
\begin{equation}
    \sum_{\mathrm{spin}}[\bar u(p)\gamma^\nu\slashed k\,v(q)]
    [\bar u(p)\gamma^\beta\slashed k\,v(q)]^*
    =8D_q\left(p^\nu k^\beta+p^\beta k^\nu-D_p\eta^{\nu\beta}\right).
\label{eq:fermion_vector_bilinear_spin_sum}
\end{equation}
Terms proportional to $k^\mu$ or $k^\nu$ vanish after contraction with the physical graviton projector, so within that contraction the last two expressions reduce to
\begin{equation}
    \sum_{\mathrm{spin}}
    [\bar u(p)v(q)]
    [\bar u(p)\gamma^\beta\slashed k\,v(q)]^*
    \longrightarrow
    4(D_qp^\beta-D_pq^\beta),
\end{equation}
and
\begin{equation}
    \sum_{\mathrm{spin}}
    [\bar u(p)\gamma^\nu\slashed k\,v(q)]
    [\bar u(p)\gamma^\beta\slashed k\,v(q)]^*
    \longrightarrow
    -8D_pD_q\,\eta^{\nu\beta}.
\end{equation}
The remaining scalar combinations are
\begin{equation}
    \Delta_\psi
    \equiv2(p\cdot q-m_\psi^2)
    =\chi-4m_\psi^2-2(D_p+D_q)
    =m_\phi^2-4m_\psi^2-2D_l,
\label{eq:app_fermion_delta_definition}
\end{equation}
and
\begin{equation}
    \Sigma_\psi
    \equiv\Delta_\psi+D_p+D_q
    =\chi-4m_\psi^2-(D_p+D_q)
    =\frac{\chi+m_\phi^2}{2}-4m_\psi^2-D_l.
\label{eq:app_fermion_sigma_definition}
\end{equation}
Using these spin sums and the transverse structures in Eq.~\eqref{eq:app_t123_definition}, the full spin- and polarization-summed result is
\begin{equation}
\label{eq:app_fermion_amplitude_squared_t123_background_mode}
\begin{aligned}
\overline{|\mathcal M_{\psi\bar\psi h}|^2}
={}&\frac{\lambda_\psi^2}{M_{\mathrm{pl}}^2}
\Bigg\{\frac{\Delta_\psi\mathcal T_1^2}{D_l^2}\\
&+\frac{\mathcal T_2}{D_p^2}
\Bigl[(\chi-4m_\psi^2)\mathcal T_2-2D_p(\mathcal T_3+D_q)\Bigr]\\
&+\frac{\mathcal T_1+\mathcal T_2-2\mathcal T_3}{D_q^2}
\Bigl[(\chi-4m_\psi^2)(\mathcal T_1+\mathcal T_2-2\mathcal T_3)
-2D_q(\mathcal T_1-\mathcal T_3+D_p)\Bigr]\\
&+\frac{-4\Sigma_\psi(\mathcal T_3^2-\tfrac12\mathcal T_1\mathcal T_2)
+2D_p\mathcal T_1\mathcal T_3}{D_lD_p}\\
&+\frac{1}{D_lD_q}\Bigl\{-4\Sigma_\psi
\Bigl[(\mathcal T_1-\mathcal T_3)^2
-\tfrac12\mathcal T_1(\mathcal T_1+\mathcal T_2-2\mathcal T_3)\Bigr]
+2D_q\mathcal T_1(\mathcal T_1-\mathcal T_3)\Bigr\}\\
&+\frac{1}{D_pD_q}\Bigg\{4\Sigma_\psi
\Bigl[(\mathcal T_3-\mathcal T_2)^2
-\tfrac12\mathcal T_2(\mathcal T_1+\mathcal T_2-2\mathcal T_3)\Bigr]\\
&\hspace{3.2em}-2(\mathcal T_3-\mathcal T_2)
\Bigl[\mathcal T_2D_q+(\mathcal T_1+\mathcal T_2-2\mathcal T_3)D_p
-2D_pD_q\Bigr]\Bigg\}\Bigg\}.
\end{aligned}
\end{equation}

\subsection{Phase-space reduction}\label{app:three_body_phase_space}

The phase space for three-body production from the collision background,
\begin{equation}
    \phi(l) \to X(p)\,\bar X(q)\,h_{\mu\nu}(k),
\end{equation}
in the collision frame, with the background momentum $\boldsymbol{\ell}$ along the $z$ direction,
\begin{equation}
    l^\mu=(E,\boldsymbol{\ell})=(E,0,0,\ell),
    \qquad
    \chi\equiv l_\mu l^\mu=E^2-\ell^2\neq m_\phi^2.
\end{equation}

The Lorentz-invariant three-body phase space is
\begin{equation}
\label{eq:app_three_body_lorentz_phase_space}
    d\Pi_{X\bar X h}
    =
    (2\pi)^4
    \delta^{(4)}(l-p-q-k)
    \frac{d^3\boldsymbol{p}}{(2\pi)^3\,2E_p}
    \frac{d^3\boldsymbol{q}}{(2\pi)^3\,2E_q}
    \frac{d^3\boldsymbol{k}}{(2\pi)^3\,2k},
\end{equation}
where the graviton is massless and the two matter particles have equal mass $m_X$,
\begin{equation}
    p_\mu p^\mu=q_\mu q^\mu=m_X^2,
    \qquad
    k_\mu k^\mu=0.
\end{equation}

Using the spatial momentum delta function $\delta^{(3)}(\boldsymbol{\ell}-\boldsymbol{p}-\boldsymbol{q}-\boldsymbol{k})$ to integrate over $\boldsymbol{q}$ gives
\begin{equation}
    \boldsymbol{q}=\boldsymbol{\ell}-\boldsymbol{p}-\boldsymbol{k},
    \qquad
    E_q=
    \sqrt{
    m_X^2+
    |\boldsymbol{\ell}-\boldsymbol{p}-\boldsymbol{k}|^2
    }.
\end{equation}
Therefore,
\begin{equation}
\label{eq:app_three_body_phase_space_after_q_integration}
    d\Pi_{X\bar X h}
    =
    \frac{
    d^3\boldsymbol{p}\,d^3\boldsymbol{k}
    }{
    8(2\pi)^5 E_pE_q k
    }
    \delta(E-E_p-E_q-k).
\end{equation}

The momentum measures satisfy
\begin{equation}
    d^3\boldsymbol k=k^2\,dk\,d\Omega_k,\qquad d^3\boldsymbol p=p^2\,dp\,d\Omega_p=pE_p\,dE_p\,d\Omega_p,\qquad p\equiv|\boldsymbol p|=\sqrt{E_p^2-m_X^2}.
\end{equation}
so that the phase space measure before carrying out the angular integrations is
\begin{equation}
    d\Pi_{X\bar Xh}=\frac{pk}{8(2\pi)^5}\,dk\,dE_p\,d\Omega_k\,d\Omega_p\,\frac{\delta(E-E_p-E_q-k)}{E_q}.
\label{eq:app_three_body_phase_space_before_angular_reduction}
\end{equation}

We parameterize the graviton and matter momenta relative to the $z$ axis along $\boldsymbol{\ell}$, keeping the azimuthal angles $\theta_k$ and $\theta_p$ independent:
\begin{equation}
\begin{aligned}
    k^\mu
    &=(k,\boldsymbol{k})
    =k
    \left(
    1,\,
    \sigma_k\cos\theta_k,\,
    \sigma_k\sin\theta_k,\,
    \mu_k
    \right),
    \\
    p^\mu
    &=(E_p,\boldsymbol{p})
    =\left(
    E_p,\,
    p\sigma_p\cos\theta_p,\,
    p\sigma_p\sin\theta_p,\,
    p\mu_p
    \right),
\end{aligned}
\end{equation}
where
\begin{equation}
    \mu_k \equiv
    \widehat{\boldsymbol{\ell}}\cdot\widehat{\boldsymbol{k}},
    \qquad
    \sigma_k \equiv \sqrt{1-\mu_k^2},
\end{equation}
and
\begin{equation}
    \mu_p \equiv
    \widehat{\boldsymbol{\ell}}\cdot\widehat{\boldsymbol{p}},
    \qquad
    \sigma_p \equiv \sqrt{1-\mu_p^2}.
\end{equation}

The azimuthal dependence is carried by the relative angle
\begin{equation}
    \Delta\theta\equiv\theta_p-\theta_k
\end{equation}
so that the angle between $\boldsymbol{p}$ and $\boldsymbol{k}$ satisfies
\begin{equation}
\label{eq:app_three_body_momentum_opening_angle}
    \zeta
    \equiv
    \widehat{\boldsymbol{p}}\cdot\widehat{\boldsymbol{k}}
    =
    \mu_p\mu_k+\sigma_p\sigma_k\cos\Delta\theta.
\end{equation}

The planar collision background is axially symmetric about $\boldsymbol{\ell}$. A common rotation of all final state momenta about the $z$ axis therefore leaves the integrand unchanged, and the angular measure can be written as
\begin{equation}
    d\Omega_k\,d\Omega_p=d\mu_k\,d\mu_p\,d\theta_k\,d\theta_p=d\mu_k\,d\mu_p\,d\theta_k\,d(\Delta\theta),\qquad \int_0^{2\pi}d\theta_k=2\pi.
\end{equation}
After integrating over the common azimuthal angle, only the relative angle $\Delta\theta$ remains. Eq.~\eqref{eq:app_three_body_phase_space_before_angular_reduction} becomes
\begin{equation}
    d\Pi_{X\bar Xh}=\frac{pk}{8(2\pi)^4}\,dk\,dE_p\,d\mu_k\,d\mu_p\,d(\Delta\theta)\,\frac{\delta(E-E_p-E_q-k)}{E_q}.
\label{eq:app_three_body_phase_space_relative_azimuth}
\end{equation}

In this parameterization,
\begin{equation}
\begin{aligned}
E_q^2={}&m_X^2+\ell^2+p^2+k^2
-2\ell p\mu_p-2\ell k\mu_k+2pk\zeta\\
={}&m_X^2+\ell^2+p^2+k^2
-2\ell p\mu_p-2\ell k\mu_k\\
&+2pk\mu_p\mu_k+2pk\sigma_p\sigma_k\cos\Delta\theta.
\end{aligned}
\label{eq:app_three_body_recoil_energy}
\end{equation}

The energy delta function fixes the relative azimuthal angle. On its support,
\begin{equation}
    E_q=E_q^\ast,
    \qquad
    E_q^\ast\equiv E-E_p-k.
\end{equation}
The corresponding energy-conserving value of $\zeta=\widehat{\boldsymbol{p}}\cdot\widehat{\boldsymbol{k}}$ is
\begin{equation}
\label{eq:app_three_body_energy_conserving_opening_angle}
    \zeta_0
    =
    \frac{
    (E-E_p-k)^2
    -m_X^2-\ell^2-p^2-k^2
    +2\ell p\,\mu_p
    +2\ell k\,\mu_k
    }{
    2pk
    }.
\end{equation}
The root for the relative azimuthal angle is therefore
\begin{equation}
\label{eq:app_three_body_relative_azimuth_root}
    u_0
    \equiv\cos(\Delta\theta_0)
    =
    \frac{
    \zeta_0-\mu_p\mu_k
    }{
    \sigma_p\sigma_k
    }.
\end{equation}
Equivalently,
\begin{equation}
    u_0=\frac{(E-E_p-k)^2-m_X^2-\ell^2-p^2-k^2+2\ell p\mu_p+2\ell k\mu_k-2pk\mu_p\mu_k}{2pk\sigma_p\sigma_k}.
\label{eq:app_three_body_relative_azimuth_root_explicit}
\end{equation}

A physical solution requires
\begin{equation}
    E_q^\ast\geq m_X,
    \qquad
    |u_0|\leq 1.
\end{equation}
For $-1<u_0<1$, there are two solutions in the interval $0\leq\Delta\theta<2\pi$,
\begin{equation}
    \Delta\theta_+=\arccos u_0,
    \qquad
    \Delta\theta_-=2\pi-\arccos u_0.
\end{equation}

Differentiating the argument of the energy delta function gives the Jacobian,
\begin{equation}
\begin{aligned}
&\left|\frac{\partial}{\partial(\Delta\theta)}
[E-E_p-k-E_q(\Delta\theta)]\right|_{\Delta\theta=\Delta\theta_\pm}\\
&\qquad=\left|\frac{pk\sigma_p\sigma_k\sin\Delta\theta}
{E_q(\Delta\theta)}\right|_{\Delta\theta=\Delta\theta_\pm}\\
&\qquad=\frac{pk\sigma_p\sigma_k}{E_q^\ast}\sqrt{1-u_0^2}.
\end{aligned}
\label{eq:app_three_body_azimuthal_delta_jacobian}
\end{equation}
Hence the azimuthal part of the phase space measure reduces to
\begin{equation}
    \int_0^{2\pi}d(\Delta\theta)\,\frac{\delta(E-E_p-E_q-k)}{E_q}=\frac{2\,\Theta(E-E_p-k-m_X)\Theta(1-u_0^2)}{pk\sigma_p\sigma_k\sqrt{1-u_0^2}}.
\label{eq:app_three_body_azimuthal_delta_reduction}
\end{equation}

The invariant mass of the two matter particles is
\begin{equation}
    s_{X\bar X}\equiv(l-k)_\mu(l-k)^\mu=\chi-2l\cdot k=\chi-2k(E-\ell\mu_k).
\label{eq:app_three_body_pair_invariant_mass}
\end{equation}
The pair production threshold requires
\begin{equation}
    s_{X\bar X}\geq 4m_X^2.
\end{equation}
At fixed $\mu_k$, this condition gives the graviton energy range
\begin{equation}
\label{eq:app_three_body_graviton_energy_bound}
    0<k\leq
    k_{\max}(\mu_k)
    =
    \frac{
    \chi-4m_X^2
    }{
    2(E-\ell\mu_k)
    },
\end{equation}
provided $\chi>4m_X^2$.

Substituting Eq.~\eqref{eq:app_three_body_azimuthal_delta_reduction} into Eq.~\eqref{eq:app_three_body_phase_space_relative_azimuth} cancels the factors of $p$ and $k$. The reduced phase space is
\begin{equation}
\label{eq:app_three_body_reduced_phase_space}
\begin{aligned}
    \frac{d\Pi_{X\bar X h}}{dk}
    =
    {}&
    \frac{1}{4(2\pi)^4}
    \int_{-1}^{1}d\mu_k\,
    \Theta(s_{X\bar X}-4m_X^2)
    \\
    &\times
    \int_{-1}^{1}d\mu_p
    \int_{m_X}^{E-k-m_X}dE_p\,
    \frac{
    \Theta(1-u_0^2)
    }{
    \sigma_k\sigma_p\sqrt{1-u_0^2}
    }.
\end{aligned}
\end{equation}
The $E_p$ integral is understood to vanish when $E-k<2m_X$. Equivalently, one may include an additional factor $\Theta(E-k-2m_X)$ in Eq.~\eqref{eq:app_three_body_reduced_phase_space}.

We then apply the optical theorem,
\begin{equation}
\label{eq:app_three_body_optical_theorem}
    \mathrm{Im}
    \left[
    \widetilde{\Gamma}^{(2)}(E,\ell)
    \right]_{X\bar X h}
    =
    \frac{1}{2}
    \int d\Pi_{X\bar X h}\,
    \overline{|\mathcal M_{X\bar Xh}|^2}.
\end{equation}
The overline denotes the sum over the two graviton polarizations and, for the fermion channel, the final-state fermion spins.

After summing over final-state spins and graviton polarizations, the squared amplitudes depend on the relative azimuthal angle only through $\cos\Delta\theta$. Thus,
\begin{equation}
    \overline{|\mathcal M_{X\bar Xh}(\Delta\theta_+)|^2}
    =
    \overline{|\mathcal M_{X\bar Xh}(\Delta\theta_-)|^2}
    =
    \left.
    \overline{|\mathcal M_{X\bar Xh}|^2}
    \right|_{\cos\Delta\theta=u_0}.
\end{equation}
The weighted version of the azimuthal reduction is thus
\begin{equation}
\begin{aligned}
&\int_0^{2\pi}d(\Delta\theta)\,
\frac{\delta(E-E_p-E_q-k)}{E_q}
\overline{|\mathcal M_{X\bar Xh}|^2}\\
&\quad=\frac{2}{pk\sigma_p\sigma_k\sqrt{1-u_0^2}}
\left.\overline{|\mathcal M_{X\bar Xh}|^2}\right|_{\cos\Delta\theta=u_0}\\
&\qquad\times\Theta(E-E_p-k-m_X)\Theta(1-u_0^2).
\end{aligned}
\label{eq:app_three_body_weighted_azimuthal_reduction}
\end{equation}
Substituting this result into Eq.~\eqref{eq:app_three_body_optical_theorem} gives the differential imaginary part,
\begin{equation}
\begin{aligned}
\frac{d}{dk}\mathrm{Im}
[\widetilde\Gamma^{(2)}(E,\ell)]_{X\bar Xh}
={}&\frac{1}{8(2\pi)^4}\int_{-1}^{1}d\mu_k\,
\Theta(s_{X\bar X}-4m_X^2)\\
&\times\int_{-1}^{1}d\mu_p
\int_{m_X}^{E-k-m_X}dE_p\\
&\times\frac{\left.\overline{|\mathcal M_{X\bar Xh}|^2}
\right|_{\cos\Delta\theta=u_0}}
{\sigma_k\sigma_p\sqrt{1-u_0^2}}\Theta(1-u_0^2).
\end{aligned}
\label{eq:app_three_body_differential_imaginary_self_energy}
\end{equation}

For the scalar final state,
\begin{equation}
\begin{aligned}
\frac{d}{dk}\operatorname{Im}
[\widetilde\Gamma^{(2)}(E,\ell)]_{\varphi\varphi^\ast h}
={}&\frac{1}{8(2\pi)^4}\int_{-1}^{1}d\mu_k\,
\Theta(s_{\varphi\varphi^\ast}-4m_\varphi^2)\\
&\times\int_{-1}^{1}d\mu_p
\int_{m_\varphi}^{E-k-m_\varphi}dE_p\\
&\times\frac{\left.\overline{|\mathcal M_{\varphi\varphi^\ast h}|^2}
\right|_{\cos\Delta\theta=u_0}}
{\sigma_k\sigma_p\sqrt{1-u_0^2}}\Theta(1-u_0^2).
\end{aligned}
\label{eq:app_scalar_three_body_differential_imaginary_self_energy}
\end{equation}

After eliminating $\Delta\theta$ with the energy delta function, the scalar products entering the squared amplitudes are evaluated at the energy-conserving solution,
\begin{equation}
    \widehat{\boldsymbol p}\!\cdot\!\widehat{\boldsymbol k}
    \to \zeta_0,
    \qquad
    \cos\Delta\theta\to u_0.
    \label{eq:app_three_body_energy_conserving_substitution}
\end{equation}
The transverse contractions and propagator denominators then follow directly from Eqs.~\eqref{eq:app_t123_definition} and \eqref{eq:app_three_body_denominators}, without introducing separate quantities with a subscript \(0\).

\subsection{Number and GW power spectra}\label{app:three_body_spectra}

Using the reduced three-body phase space, the graviton number spectrum per unit wall area is
\begin{equation}
\begin{aligned}
\frac{dN_{X\bar Xh}}{A\,dk}
={}&\frac{v_\phi^2}{4\pi^6}
\int_{\Lambda_{\min}}^{\Lambda_{\max}}dE
\int_{\Lambda_{\min}}^{\Lambda_{\max}}d\ell\,
\frac{\Theta(\chi)}{\chi^2}\\
&\times\int_{-1}^{1}d\mu_k\,
\Theta(s_{X\bar X}-4m_X^2)
\int_{-1}^{1}d\mu_p\\
&\times\int_{m_X}^{E-k-m_X}dE_p\,
\frac{\left.\overline{|\mathcal M_{X\bar Xh}|^2}
\right|_{\cos\Delta\theta=u_0}}
{\sigma_k\sigma_p\sqrt{1-u_0^2}}\Theta(1-u_0^2).
\end{aligned}
\label{eq:app_three_body_differential_graviton_yield}
\end{equation}
Here the kinematic quantities are those defined in Eqs.~\eqref{eq:app_three_body_four_momenta}--\eqref{eq:app_three_body_denominators} and in the phase-space reduction above. The scalar and fermion squared amplitudes are given in Eqs.~\eqref{eq:scalar_amplitude_squared_background_mode} and \eqref{eq:app_fermion_amplitude_squared_t123_background_mode}, respectively.

For numerical integration, we exchange the integration order and make the limits explicit. At fixed \(k\), \(\ell\), and \(\mu_k\), the pair threshold \(s_{X\bar X}\geq 4m_X^2\) gives
\begin{equation}
    E\geq E_-(\ell,k,\mu_k),\qquad E_-(\ell,k,\mu_k)=k+\sqrt{\ell^2+k^2-2\ell k\mu_k+4m_X^2}.
\label{eq:app_three_body_minimum_background_energy}
\end{equation}
The triangle inequality gives \(E_-(\ell,k,\mu_k)\geq\ell\). Since the background window requires \(\ell\geq\Lambda_{\min}\), the lower cutoff on \(E\) is automatically satisfied by this kinematic limit.
Together with the cutoff \(E\leq\Lambda_{\max}\), this requires
\begin{equation}
    \ell^2+k^2-2\ell k\mu_k\leq(\Lambda_{\max}-k)^2-4m_X^2.
\end{equation}
Energy conservation therefore gives the necessary upper bound
\begin{equation}
    0<k<\Lambda_{\max}-2m_X.
    \label{eq:app_three_body_graviton_endpoint}
\end{equation}
For a given \(k\), define
\begin{equation}
    Q(k)\equiv\sqrt{(\Lambda_{\max}-k)^2-4m_X^2}.
\end{equation}
The allowed range of \(\ell\) at fixed \(k\) and \(\mu_k\) is obtained by intersecting \(\Lambda_{\min}\leq\ell\leq\Lambda_{\max}\) with
\begin{equation}
    \ell^2-2\ell k\mu_k+k^2\leq Q^2(k).
\end{equation}
Therefore,
\begin{equation}
    \ell_-(k,\mu_k)
    =
    \max\!\left\{\Lambda_{\min},\,k\mu_k-
    \sqrt{Q^2(k)-k^2(1-\mu_k^2)}\right\},
\end{equation}
\begin{equation}
    \ell_+(k,\mu_k)
    =
    \min\!\left\{\Lambda_{\max},\,k\mu_k+
    \sqrt{Q^2(k)-k^2(1-\mu_k^2)}\right\},
\end{equation}
with the additional requirement \(\ell_+>\ell_-\) and the implicit condition
\begin{equation}
    Q^2(k)-k^2(1-\mu_k^2)\geq 0.
\end{equation}
The three-body graviton number spectrum then becomes
\begin{equation}
\begin{aligned}
\frac{dN_{X\bar Xh}}{A\,dk}
={}&\frac{v_\phi^2}{4\pi^6}
\int_{-1}^{1}d\mu_k
\int_{\ell_-(k,\mu_k)}^{\ell_+(k,\mu_k)}d\ell\\
&\times\int_{E_-(\ell,k,\mu_k)}^{\Lambda_{\max}}dE\,
\frac{1}{\chi^2}
\int_{-1}^{1}d\mu_p\\
&\times\int_{m_X}^{E-k-m_X}dE_p\,
\frac{\left.\overline{|\mathcal M_{X\bar Xh}|^2}
\right|_{\cos\Delta\theta=u_0}}
{\sigma_k\sigma_p\sqrt{1-u_0^2}}\Theta(1-u_0^2).
\end{aligned}
\label{eq:app_three_body_explicit_limit_graviton_yield}
\end{equation}
This expression is the three-body counterpart of the two-body spectrum with explicit integration limits. Substituting the scalar or fermion squared amplitude derived in Appendix~\ref{app:three_body_calculation} gives the corresponding differential graviton spectrum.

The corresponding energy spectrum per logarithmic interval is
\begin{equation}
    \frac{d\mathcal E_{X\bar Xh}}{A\,d\ln k}
    =
    k^2\frac{dN_{X\bar Xh}}{A\,dk}.
\end{equation}
Multiplying by the geometric factor \(3/(2R_*)\) converts the number per unit area to a volume density. The GW spectrum at production is then
\begin{equation}
\label{eq:app_three_body_gw_energy_spectrum}
    \Omega_{\mathrm{GW}}^{*(X\bar Xh)}(k)
    =
    \frac{3}{2R_*\rho_{\mathrm{tot}}}
    k^2
    \frac{dN_{X\bar Xh}}{A\,dk}.
\end{equation}
Using the redshift prescription in Eq.~\eqref{eq:gw_redshift_to_present_epoch}, the present-day spectrum is
\begin{equation}
    h^2\Omega_{{\mathrm{GW}},0}^{(X\bar Xh)}(k_0)
    \simeq
    1.7\times10^{-5}
    \left(\frac{100}{g_*}\right)^{1/3}
    \Omega_{\mathrm{GW}}^{*(X\bar Xh)}(k),
\end{equation}
with
\begin{equation}
    k_0=\frac{T_0}{T_*}\left(\frac{g_{*s,0}}{g_*}\right)^{1/3}k.
\end{equation}

\bibliography{ref}

@article{Watkins:1991zt,
    author = "Watkins, Richard and Widrow, Lawrence M.",
    title = "{Aspects of Reheating in First Order Inflation}",
    reportNumber = "FERMILAB-PUB-91-164-A",
    doi = "10.1016/0550-3213(92)90362-F",
    journal = "Nucl. Phys. B",
    volume = "374",
    pages = "446--468",
    year = "1992"
}

@article{Falkowski:2012fb,
    author = "Falkowski, Adam and No, Jose M.",
    title = "{Non-thermal Dark Matter Production from the Electroweak Phase Transition: Multi-TeV WIMPs and 'Baby-Zillas'}",
    eprint = "1211.5615",
    archivePrefix = "arXiv",
    primaryClass = "hep-ph",
    reportNumber = "ULB-TH-12-18, LPT-12-113",
    doi = "10.1007/JHEP02(2013)034",
    journal = "JHEP",
    volume = "02",
    pages = "034",
    year = "2013"
}

@article{Mansour:2023fwj,
    author = "Mansour, Henda and Shakya, Bibhushan",
    title = "{Particle production from phase transition bubbles}",
    eprint = "2308.13070",
    archivePrefix = "arXiv",
    primaryClass = "hep-ph",
    reportNumber = "DESY-23-121, TTP23-033, P3H-23-056",
    doi = "10.1103/PhysRevD.111.023520",
    journal = "Phys. Rev. D",
    volume = "111",
    number = "2",
    pages = "023520",
    year = "2025"
}

@article{Barman:2023ymn,
    author = "Barman, Basabendu and Bernal, Nicol{\'a}s and Xu, Yong and Zapata, {\'O}scar",
    title = "{Gravitational wave from graviton Bremsstrahlung during reheating}",
    eprint = "2301.11345",
    archivePrefix = "arXiv",
    primaryClass = "hep-ph",
    doi = "10.1088/1475-7516/2023/05/019",
    journal = "JCAP",
    volume = "05",
    pages = "019",
    year = "2023"
}

@article{Gleisberg:2003,
  author        = {Gleisberg, Tanju and Krauss, Frank and Matchev, Konstantin T. and Schalicke, Andreas and Schumann, Steffen and Soff, Gerhard},
  title         = {Helicity Formalism for Spin-2 Particles},
  journal       = {Journal of High Energy Physics},
  volume        = {2003},
  number        = {09},
  pages         = {001},
  year          = {2003},
  eprint        = {hep-ph/0306182},
  archivePrefix = {arXiv},
  doi           = {10.1088/1126-6708/2003/09/001}
}

@article{Shakya:2023pp,
  author        = {Shakya, Bibhushan},
  title         = {Aspects of Particle Production from Bubble Dynamics at a First Order Phase Transition},
  journal       = {Phys. Rev. D},
  volume        = {111},
  number        = {2},
  pages         = {023521},
  year          = {2025},
  eprint        = {2308.16224},
  archivePrefix = {arXiv},
  primaryClass  = {hep-ph},
  doi           = {10.1103/PhysRevD.111.023521}
}

@article{Inomata:2024rkt,
    author = "Inomata, Keisuke and Kamionkowski, Marc and Kasai, Kentaro and Shakya, Bibhushan",
    title = "{Gravitational waves from particles produced from bubble collisions in first-order phase transitions}",
    eprint = "2412.17912",
    archivePrefix = "arXiv",
    primaryClass = "astro-ph.CO",
    doi = "10.1103/k4s5-8zqy",
    journal = "Phys. Rev. D",
    volume = "112",
    number = "8",
    pages = "083523",
    year = "2025"
}

@article{Giudice:2024pp,
  author        = {Giudice, Gian F. and Lee, Hyun Min and Pomarol, Alex and Shakya, Bibhushan},
  title         = {Nonthermal Heavy Dark Matter from a First-Order Phase Transition},
  journal       = {JHEP},
  volume        = {12},
  pages         = {190},
  year          = {2024},
  eprint        = {2403.03252},
  archivePrefix = {arXiv},
  primaryClass  = {hep-ph},
  doi           = {10.1007/JHEP12(2024)190}
}

@article{Nakayama:2018pt,
  author  = {Nakayama, Kazunori and Tang, Yong},
  title   = {Stochastic Gravitational Waves from Particle Origin},
  journal = {Phys. Lett. B},
  volume  = {788},
  pages   = {341--346},
  year    = {2019},
  doi     = {10.1016/j.physletb.2018.11.023}
}

@article{Herman:2023hfgw,
  author  = {Herman, Nicolas and Lehoucq, L{\'e}onard and F{\'u}zfa, Andr{\'e}},
  title   = {Electromagnetic Antennas for the Resonant Detection of the Stochastic Gravitational Wave Background},
  journal = {Phys. Rev. D},
  volume  = {108},
  number  = {12},
  pages   = {124009},
  year    = {2023},
  doi     = {10.1103/PhysRevD.108.124009}
}

@article{Kosowsky:1991ua,
  author  = {Kosowsky, Arthur and Turner, Michael S. and Watkins, Richard},
  title   = {Gravitational Radiation from Colliding Vacuum Bubbles},
  journal = {Phys. Rev. D},
  volume  = {45},
  pages   = {4514--4535},
  year    = {1992},
  doi     = {10.1103/PhysRevD.45.4514}
}

@article{Kosowsky:1992vn,
  author        = {Kosowsky, Arthur and Turner, Michael S.},
  title         = {Gravitational Radiation from Colliding Vacuum Bubbles: Envelope Approximation to Many-Bubble Collisions},
  journal       = {Phys. Rev. D},
  volume        = {47},
  pages         = {4372--4391},
  year          = {1993},
  eprint        = {astro-ph/9211004},
  archivePrefix = {arXiv},
  doi           = {10.1103/PhysRevD.47.4372}
}

@article{Kamionkowski:1993fg,
  author        = {Kamionkowski, Marc and Kosowsky, Arthur and Turner, Michael S.},
  title         = {Gravitational Radiation from First-Order Phase Transitions},
  journal       = {Phys. Rev. D},
  volume        = {49},
  pages         = {2837--2851},
  year          = {1994},
  eprint        = {astro-ph/9310044},
  archivePrefix = {arXiv},
  doi           = {10.1103/PhysRevD.49.2837}
}

@article{Espinosa:2010hh,
  author        = {Espinosa, Jose R. and Konstandin, Thomas and No, Jose M. and Servant, Geraldine},
  title         = {Energy Budget of Cosmological First-Order Phase Transitions},
  journal       = {JCAP},
  volume        = {06},
  pages         = {028},
  year          = {2010},
  eprint        = {1004.4187},
  archivePrefix = {arXiv},
  primaryClass  = {hep-ph},
  doi           = {10.1088/1475-7516/2010/06/028}
}

@article{Caprini:2015zlo,
  author        = {Caprini, Chiara and Hindmarsh, Mark and Huber, Stephan and Konstandin, Thomas and Kozaczuk, Jonathan and Nardini, Germano and No, Jose M. and Petiteau, Antoine and Schwaller, Pedro and Servant, Geraldine and Weir, David J.},
  title         = {Science with the Space-Based Interferometer {eLISA}. II: Gravitational Waves from Cosmological Phase Transitions},
  journal       = {JCAP},
  volume        = {04},
  pages         = {001},
  year          = {2016},
  eprint        = {1512.06239},
  archivePrefix = {arXiv},
  primaryClass  = {astro-ph.CO},
  doi           = {10.1088/1475-7516/2016/04/001}
}

@article{Caprini:2019egz,
  author        = {Caprini, Chiara and Chala, Mikael and Dorsch, Glauber C. and Hindmarsh, Mark and Huber, Stephan J. and Konstandin, Thomas and Kozaczuk, Jonathan and Nardini, Germano and No, Jose M. and Rummukainen, Kari and Schwaller, Pedro and Servant, Geraldine and Tranberg, Anders and Weir, David J.},
  title         = {Detecting Gravitational Waves from Cosmological Phase Transitions with {LISA}: An Update},
  journal       = {JCAP},
  volume        = {03},
  pages         = {024},
  year          = {2020},
  eprint        = {1910.13125},
  archivePrefix = {arXiv},
  primaryClass  = {astro-ph.CO},
  doi           = {10.1088/1475-7516/2020/03/024}
}

@article{Hindmarsh:2013xza,
  author        = {Hindmarsh, Mark and Huber, Stephan J. and Rummukainen, Kari and Weir, David J.},
  title         = {Gravitational Waves from the Sound of a First Order Phase Transition},
  journal       = {Phys. Rev. Lett.},
  volume        = {112},
  pages         = {041301},
  year          = {2014},
  eprint        = {1304.2433},
  archivePrefix = {arXiv},
  primaryClass  = {hep-ph},
  doi           = {10.1103/PhysRevLett.112.041301}
}

@article{Hindmarsh:2015qta,
  author        = {Hindmarsh, Mark and Huber, Stephan J. and Rummukainen, Kari and Weir, David J.},
  title         = {Numerical Simulations of Acoustically Generated Gravitational Waves at a First Order Phase Transition},
  journal       = {Phys. Rev. D},
  volume        = {92},
  number        = {12},
  pages         = {123009},
  year          = {2015},
  eprint        = {1504.03291},
  archivePrefix = {arXiv},
  primaryClass  = {astro-ph.CO},
  doi           = {10.1103/PhysRevD.92.123009}
}

@article{Hindmarsh:2017gnf,
  author        = {Hindmarsh, Mark and Huber, Stephan J. and Rummukainen, Kari and Weir, David J.},
  title         = {Shape of the Acoustic Gravitational Wave Power Spectrum from a First Order Phase Transition},
  journal       = {Phys. Rev. D},
  volume        = {96},
  number        = {10},
  pages         = {103520},
  year          = {2017},
  eprint        = {1704.05871},
  archivePrefix = {arXiv},
  primaryClass  = {astro-ph.CO},
  doi           = {10.1103/PhysRevD.96.103520},
  note          = {[Erratum: Phys. Rev. D 101, 089902 (2020)]}
}

@article{RoperPol:2019wvy,
  author        = {Roper Pol, Alberto and Mandal, Sayan and Brandenburg, Axel and Kahniashvili, Tina and Kosowsky, Arthur},
  title         = {Numerical Simulations of Gravitational Waves from Early-Universe Turbulence},
  journal       = {Phys. Rev. D},
  volume        = {102},
  number        = {8},
  pages         = {083512},
  year          = {2020},
  eprint        = {1903.08585},
  archivePrefix = {arXiv},
  primaryClass  = {astro-ph.CO},
  doi           = {10.1103/PhysRevD.102.083512}
}

@article{Bodeker:2009qy,
  author        = {B{\"o}deker, Dietrich and Moore, Guy D.},
  title         = {Can Electroweak Bubble Walls Run Away?},
  journal       = {JCAP},
  volume        = {05},
  pages         = {009},
  year          = {2009},
  eprint        = {0903.4099},
  archivePrefix = {arXiv},
  primaryClass  = {hep-ph},
  doi           = {10.1088/1475-7516/2009/05/009}
}

@article{Bodeker:2017cim,
  author        = {B{\"o}deker, Dietrich and Moore, Guy D.},
  title         = {Electroweak Bubble Wall Speed Limit},
  journal       = {JCAP},
  volume        = {05},
  pages         = {025},
  year          = {2017},
  eprint        = {1703.08215},
  archivePrefix = {arXiv},
  primaryClass  = {hep-ph},
  doi           = {10.1088/1475-7516/2017/05/025}
}

@article{Lewicki:2019gmv,
  author        = {Lewicki, Marek and Vaskonen, Ville},
  title         = {On Bubble Collisions in Strongly Supercooled Phase Transitions},
  journal       = {Phys. Dark Univ.},
  volume        = {30},
  pages         = {100672},
  year          = {2020},
  eprint        = {1912.00997},
  archivePrefix = {arXiv},
  primaryClass  = {astro-ph.CO},
  doi           = {10.1016/j.dark.2020.100672}
}

@article{Barman:2023monomial,
  author        = {Barman, Basabendu and Bernal, Nicol{\'a}s and Xu, Yong and Zapata, {\'O}scar},
  title         = {Bremsstrahlung-Induced Gravitational Waves in Monomial Potentials during Reheating},
  journal       = {Phys. Rev. D},
  volume        = {108},
  number        = {8},
  pages         = {083524},
  year          = {2023},
  eprint        = {2305.16388},
  archivePrefix = {arXiv},
  primaryClass  = {hep-ph},
  doi           = {10.1103/PhysRevD.108.083524}
}

@article{Aggarwal:2020olq,
  author        = {Aggarwal, Nancy and others},
  title         = {Challenges and Opportunities of Gravitational-Wave Searches at {MHz} to {GHz} Frequencies},
  journal       = {Living Rev. Rel.},
  volume        = {24},
  number        = {1},
  pages         = {4},
  year          = {2021},
  eprint        = {2011.12414},
  archivePrefix = {arXiv},
  primaryClass  = {gr-qc},
  doi           = {10.1007/s41114-021-00032-5}
}

@article{Jinno:2022fom,
    author = {Jinno, Ryusuke and Shakya, Bibhushan and van de Vis, Jorinde},
    title = {Gravitational Waves from Feebly Interacting Particles in a First Order Phase Transition},
    eprint = {2211.06405},
    archivePrefix = {arXiv},
    primaryClass = {gr-qc},
    reportNumber = {DESY-22-172, IFT-UAM/CSIC-22-140, MITP-22-095, RESCEU-22/22},
    doi = {10.1103/phwp-jsvq},
    journal = {Phys. Rev. Lett.},
    volume = {136},
    number = {13},
    pages = {131002},
    year = {2026}
}

@article{Ai:2025fqw,
    author = {Ai, Wen-Yuan},
    title = {High-Frequency Gravitational Waves from First-Order Phase Transitions},
    eprint = {2508.02794},
    archivePrefix = {arXiv},
    primaryClass = {hep-ph},
    doi = {10.1103/4v2z-fqsx},
    journal = {Phys. Rev. D},
    volume = {113},
    number = {5},
    pages = {056007},
    year = {2026}
}

@article{Qiu:2025tmn,
    author = {Qiu, Dayun and Jiang, Siyu and Huang, Fa Peng},
    title = {A New Source of Phase Transition Gravitational Waves: Heavy Particle Braking across Bubble Walls},
    eprint = {2508.04314},
    archivePrefix = {arXiv},
    primaryClass = {hep-ph},
    doi = {10.1007/JHEP02(2026)112},
    journal = {JHEP},
    volume = {02},
    pages = {112},
    year = {2026}
}

@article{Cheng:2026npt,
    author = "Cheng, Zihong and Huang, Fa Peng",
    title = {Dark Matter Production from Bubble Collisions during a First-Order Phase Transition at the End of Inflation},
    eprint = "2605.03758",
    archivePrefix = "arXiv",
    primaryClass = "hep-ph",
    month = "5",
    year = "2026"
}

@article{Ghoshal:2026hev,
    author = "Ghoshal, Anish and Pal, Pratyay",
    title = "{Cosmic collider gravitational waves sourced by right-handed neutrino production from bubbles: Testing scales of the seesaw mechanism, leptogenesis, and dark matter}",
    eprint = "2601.02458",
    archivePrefix = "arXiv",
    primaryClass = "astro-ph.CO",
    reportNumber = "Phys. Rev. D 113, 103035",
    doi = "10.1103/1t75-24fn",
    journal = "Phys. Rev. D",
    volume = "113",
    number = "10",
    pages = "103035",
    year = "2026"
}

@article{Cataldi:2025nac,
    author = {Cataldi, Martina and M{\"u}{\"u}rsepp, Kristjan and Vanvlasselaer, Miguel},
    title = "{CP-violation in production of heavy neutrinos from bubble collisions}",
    eprint = "2506.12123",
    archivePrefix = "arXiv",
    primaryClass = "hep-ph",
    reportNumber = "DESY-25-082",
    doi = "10.1007/JHEP01(2026)058",
    journal = "JHEP",
    volume = "01",
    pages = "058",
    year = "2026"
}

@article{Ai:2024ikj,
    author = "Ai, Wen-Yuan and Fairbairn, Malcolm and Mimasu, Ken and You, Tevong",
    title = "{Non-thermal production of heavy vector dark matter from relativistic bubble walls}",
    eprint = "2406.20051",
    archivePrefix = "arXiv",
    primaryClass = "hep-ph",
    reportNumber = "KCL-PH-TH/2024-38",
    doi = "10.1007/JHEP05(2025)225",
    journal = "JHEP",
    volume = "05",
    pages = "225",
    year = "2025",
    note = "[Erratum: JHEP 05, 021 (2026)]"
}

@article{Katz:2016adq,
    author = "Katz, Andrey and Riotto, Antonio",
    title = "{Baryogenesis and Gravitational Waves from Runaway Bubble Collisions}",
    eprint = "1608.00583",
    archivePrefix = "arXiv",
    primaryClass = "hep-ph",
    reportNumber = "CERN-TH-2016-173",
    doi = "10.1088/1475-7516/2016/11/011",
    journal = "JCAP",
    volume = "11",
    pages = "011",
    year = "2016"
}

@article{Bhandari:2026uqd,
    author = "Bhandari, Dipendu and Kumar Mandal, Rajat and Sil, Arunansu",
    title = "{First Order Phase Transition Induced Graviton Bremsstrahlung: A Multi Peak Gravitational Wave Signature}",
    eprint = "2608.14782",
    archivePrefix = "arXiv",
    primaryClass = "hep-ph",
    month = "8",
    year = "2026"
}

\end{document}